 \documentclass[final,1p,times]{elsarticle}

\usepackage{amssymb}
\usepackage{amsmath}
\usepackage{acronym}
\usepackage[OT2,T1]{fontenc}
\usepackage{color}

\newcommand{\red}[1]{#1}

\newcommand{\geant}{{\sc{Geant4}}}

\newcommand{\het}{{$^{3}$He}}
\newcommand{\talys}{{\sf{TALYS}}}

\biboptions{sort&compress}
 \usepackage{lineno}

\journal{Nucl. Instrum. Methods Phys. Res. A}

\begin{document}

\begin{frontmatter}



\title{ELIGANT-TN -- ELI Gamma Above Neutron Threshold:\\ The Thermal Neutron setup}


\author[eli]{P.-A.~S\"{o}derstr\"{o}m\corref{cor1}}\ead{par.anders@eli-np.ro}
\author[eli]{D.~L.~Balabanski} 
\author[eli]{M.~Cuciuc} 
\author[ifin]{D.~M.~Filipescu} 
\author[ifin]{I.~Gheorghe} 
\author[eli,uni-istanbul]{A.~Ku\c{s}o\u{g}lu} 
\author[eli]{C.~Matei} 
\author[eli]{D.~Testov} 
\author[eli]{S.~Aogaki} 
\author[eli]{H.~T.~Aslani} 
\author[eli]{L.~Capponi\fnref{fn1}} 
\author[eli,ropar]{D.~Choudhury} 
\author[eli,ifin]{G.~Ciocan} 
\author[ifin]{T.~Glodariu\fnref{fn2}} 
\author[eli]{M.~Krzysiek\fnref{fn4}} 
\author[eli]{V.~Lelasseux} 
\author[eli,ropar]{R.~Roy} 
\author[ifin-dfna]{R.F.~Andrei}
\author[eli]{M.~Brezeanu} 
\author[eli]{R.~Corbu} 
\author[eli]{A.~Dhal} 
\author[ifin-dfna]{D.~Iancu}
\author[eli]{D.~Kahl\fnref{fn5}} 
\author[npl]{S.~Ioannidis} 
\author[ifin]{K.~KeunHwan} 
\author[npl]{G.~Lorusso} 
\author[kazak]{B.~Mauyey} 
\author[eli]{T.~Petruse} 
\author[eli]{G.V.~Turturic\u{a}} 
\cortext[cor1]{Corresponding author}
\fntext[fn1]{Present address: National Nuclear Laboratory, Sellafield, Cumbria CA20 1PG, United Kingdom}
\fntext[fn2]{Deceased}
\fntext[fn4]{Present address: Institute of Nuclear Physics Polish Academy of Sciences, PL-31342 Cracow, Poland}
\fntext[fn5]{Present address: Facility for Rare Isotope Beams, Michigan State University, 640 South Shal Lane, East Lansing, MI 48824, USA}

\address[eli]{Extreme Light Infrastructure-Nuclear Physics (ELI-NP), Horia Hulubei National Institute for Physics and Nuclear Engineering (IFIN-HH), Str. Reactorului 30, 077125 Bucharest-M\u{a}gurele, Romania}
\address[ifin]{Department of Nuclear Physics (DFN), Horia Hulubei National Institute for Physics and Nuclear Engineering (IFIN-HH), Str. Reactorului 30, 077125 Bucharest-M\u{a}gurele, Romania}
\address[uni-istanbul]{Department of Physics, Faculty of Science, Istanbul University, Vezneciler/Fatih, 34134, Istanbul, Turkey}
\address[ropar]{Department of Physics, Indian Institute of Technology, Ropar-140001, India}
\address[ifin-dfna]{Applied Nuclear Physics Department (DFNA), Horia Hulubei National Institute for Physics and Nuclear Engineering, Str. Reactorului 30, Bucharest-M\u{a}gurele 077125, Romania}
\address[npl]{National Physical Laboratory, TW11 0LW, Teddington, UK}
\address[kazak]{L.N. Gumilyov Eurasian National University, Satpayev 2, 010008, Astana, Kazakhstan}
\begin{abstract}
Here we present the thermal neutron counter from the ELI Gamma Above Neutron Threshold setup at the Extreme Light Infrastructure – Nuclear Physics. We describe the mechanical design of the setup, the properties of the ${}^{3}$He gas counters, and the hardware data-acquisition electronics and software digital signal processing. The performance of the complete detector array is demonstrated via Geant4 and MCNP simulations, and measurements with typical neutron sources. The analysis procedure for experimental measurements are outlined with a in-beam test experiment with an $\alpha$ beam to measure the ${}^{13}\mathrm{C}(\alpha,\mathrm{n}_{0}){}^{16}\mathrm{O}$ cross-section branching ratios.
\end{abstract}

\begin{keyword}

Gas counters \sep ${}^{3}$He detectors \sep Neutron detectors \sep Thermal neutrons \sep Cross-section measurements



\end{keyword}

\end{frontmatter}


\section{Introduction}

The \ac{ELI-NP} facility \cite{Filipescu2015,Gales2016,Gales2018,Tanaka2020} has recently been constructed in Romania and will provide the international nuclear physics community with an unprecedented beam-lines within two frontiers of nuclear photonics: the completed \ac{HPLS} beam lines \cite{Lureau2020} will provide laser pulses up to 10~PW of power for nuclear physics studies \cite{Negoita2016}, and the \ac{ELI-GBS} \cite{Constantin2024} under implementation will provide high-brilliance beams of $\gamma$-rays up to 20~MeV with an intensity of $\sim$5000 photons/s/eV for experiments on nuclear structure, nuclear reactions, and nuclear astrophysics.

One of the flagship setups for the \ac{ELI-GBS} is the \ac{ELIGANT} suite of instruments \cite{Camera2016,Krzysiek2019b,Soderstrom2019a,Soderstrom2022,Clisu2023,Soderstrom2024a}, aiming to measure $\gamma$ rays \cite{Camera2016,Krzysiek2019a,Soderstrom2019a,Soderstrom2020} and neutrons \cite{Camera2016,Krzysiek2019a,Utsunomiya2017,Gheorghe2021,Clisu2023} following photonuclear reactions. There are currently two main members of the \ac{ELIGANT} family, where the \ac{ELIGANT-GN} setup \cite{Camera2016,Krzysiek2019a,Soderstrom2022} will focus on \ac{GDR} studies, in particular regarding competition between neutron- and $\gamma$-ray emission channels.
Here, we will present the other member of the family, \ac{ELIGANT-TN} \cite{Camera2016,Utsunomiya2017,Gheorghe2021} setup, which is a neutron counter designed for photo-induced cross-section measurements with applications for nuclear industry, nuclear astrophysics, and nuclear medicine \cite{Filipescu2015} to be performed at the \ac{ELI-GBS} \cite{Constantin2024} with all the challenges of using a photon beam in mind \cite{Utsunomiya2018,Filipescu2023}.

\begin{figure}[ht!]
 \begin{center}
 \includegraphics[width=0.5\columnwidth]{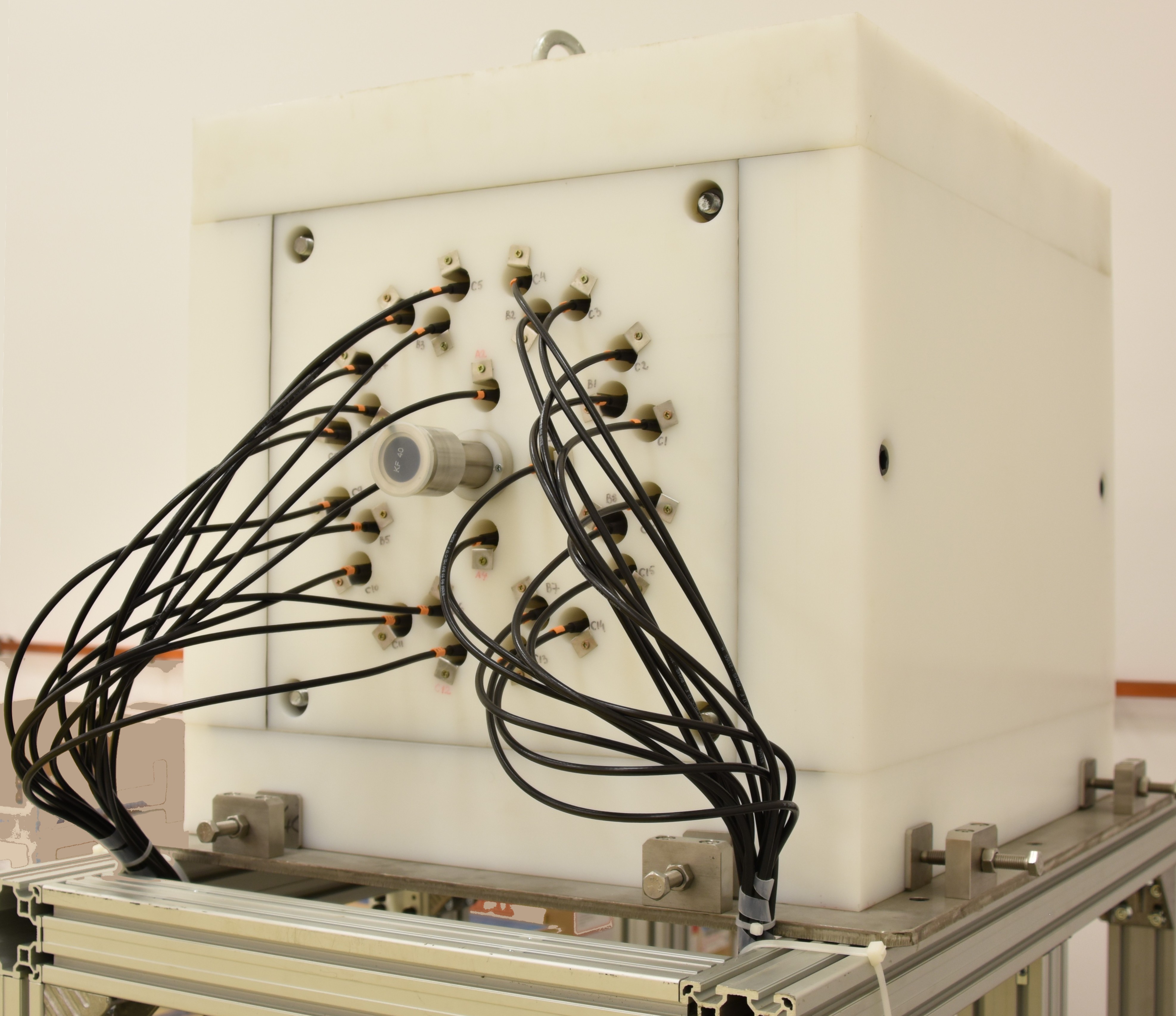}
\end{center}
\caption{Photograph of the ELIGANT-TN neutron counter.\label{fig:cube_diagonal}}
\end{figure}

\ac{ELIGANT-TN} is a neutron counter consisting of a large polyethylene moderator with 28 \het\ gas detectors, with a cross-section for thermal neutron capture about {\red{5316~b}} {\red{\cite{Carlson2009}}}
, embedded for neutron counting. Figure~\ref{fig:cube_diagonal} shows a photograph of the instrument.

The physics program of \ac{ELIGANT-TN} focuses on high-precision cross-section measurements for applications in astrophysics, industry, and medicine \cite{Camera2016,Filipescu2015,Soderstrom2023b}. The goals of this program closely follow a series of similar experiments performed at the NewSUBARU facility, serving both as an extensive stand-alone physics campaign as well as a preparatory phase for \ac{ELIGANT} experiments at \ac{ELI-NP}
\cite{Filipescu2014,Nyhus2015,Gheorghe2017a,Renstrom2018,Utsunomiya2019,Gheorghe2024}. A part of the motivation for the campaign at NewSUBARU was to address the nuclear data need for applications such as radiation shielding, radiotherapy, the development of next-generation fission and fusion reactors, safeguards, and nuclear waste management. 
This project culminated in two significant publications \cite{Goriely2019b,Kawano2020}. However, the interest in improving and expanding this data with the narrow-bandwidth beams at \ac{ELI-NP} remains. A similar setup is, furthermore, already operational at the \ac{SLEGS} facility in China \cite{Hao2025}, where an extensive program focused on resolving the discrepancy between the Saclay and \ac{LLNL} photo-neutron cross section data bases \cite{Varlamov2004} is underway. {\red{The flat-efficiency design approach is, furhtermore, of general interest within the community to minimise the systematic errors associates with the unknown neutron energies. For example, the \ac{HeBGB} detector \cite{Brandenburg2022} has been designed with a lower absolute efficiency but a very large flat-efficiency range in mind.}}

For nuclear astrophysics, there are 35 stable proton-rich isotopes, known as p-process nuclei, believed to be produced by proton capture and $\gamma$ dissociation processes in proton-rich stellar environments. The details of the p-process need to be better understood, and many p-nuclei are underproduced in astrophysical model calculations.
In this case, nuclear $(\gamma,\mathrm{n})$ reaction cross sections above the neutron separation threshold from the ground state can be obtained directly from measurements at \ac{ELI-NP}.
As the natural abundances of $p$-nuclei are very low, high intensity $\gamma$-ray beams are required for the $(\gamma,\mathrm{n})$ measurements. Some highlight reactions on this topic include $^{138}$La$(\gamma,\mathrm{n})$ which is the most underproduced nucleus in p-process calculations and $^{180}$Ta$(\gamma,\mathrm{n})$ with a very difficult to acquire target material that has made measurements unfeasible prior to \ac{ELI-NP} because of insufficient beam intensity.

As part of a complementary experimental campaign, a series of experiments has been carried out at the 3~MV Tandem facility at the \ac{IFIN-HH} \cite{Burducea2015,Velisa2021}, which also partially supports accelerator characterisation \cite{Testov2025a}. This program mainly addresses $(\alpha,\mathrm{n})$ cross-section measurements in the $\alpha$-energy range typical for actinide $\alpha$-decay, providing nuclear data input for neutron background evaluation in various applications \cite{Roy2023a,Roy2024a,Roy2024b,RoyUnp2025a}.

\section{Mechanical Design\label{sec:mechanicaldesign}}

\begin{figure}[ht!]
 \begin{center}
 \includegraphics[width=0.5\columnwidth]{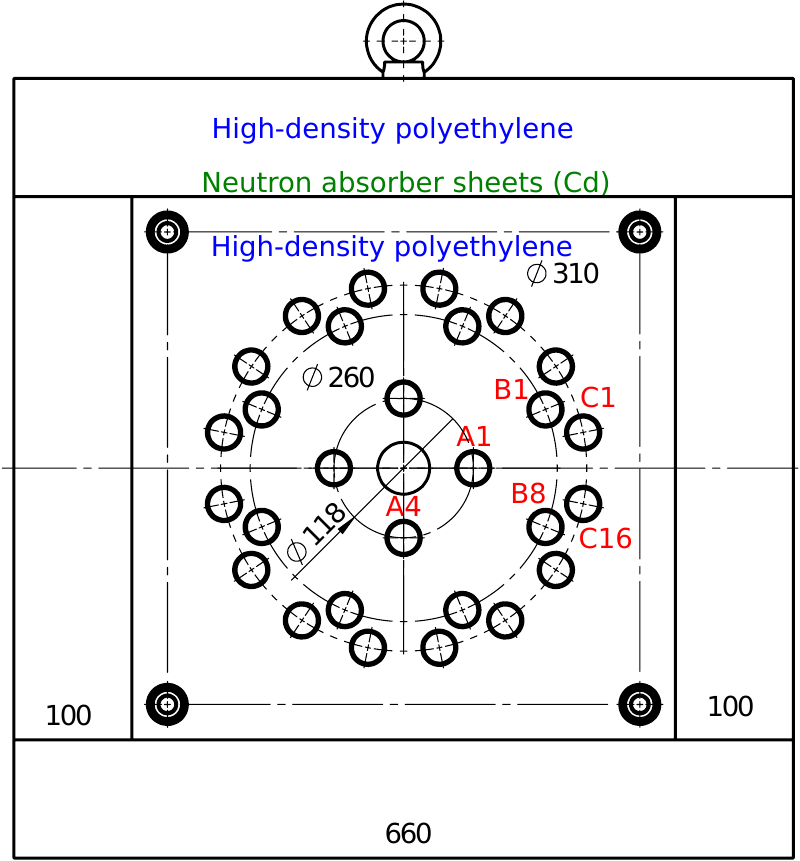}
\end{center}
\caption{Drawing of \ac{ELIGANT-TN} as viewed from the front (downstream). The size of the shielding as well as the radii of the three rings containing \het\ counters are shown in the units of mm.\label{fig:cube_front}}
\end{figure}

The main body of the moderator volume is made out of \ac{HDPE} blocks with an area of $46\times46$~cm$^{2}$, four with a thickness of 10~cm and two with a thickness of 12~cm, giving a total length of the moderator body of 64~cm. Through this body, 28 holes with a 2.6~cm diameter are drilled in a pattern of three rings, containing 4, 8, and 16 holes, respectively, as shown in Figure~\ref{fig:cube_front}. On the upstream side of \ac{ELIGANT-TN} is an additional \ac{HDPE} shielding block with a volume of $46\times46\times10$~cm$^{3}$. Through all these blocks, a 4.4~cm hole is drilled to allow for the \ac{ELI-GBS} beam to pass through and interact with the target in the centre of the array.

To shield \ac{ELIGANT-TN} from external neutrons, the sides of the moderator body are covered in thin cadmium sheets that act as an absorber for thermal neutrons. Surrounding the entire instrument is another layer of 10~cm thick \ac{HDPE} blocks to thermalise the neutrons before the cadmium shields. The final layout is shown in Figures~\ref{fig:cube_diagonal} and \ref{fig:cube_front}.

\section{\het\ detectors}

\begin{figure}[ht!]
 \begin{center}
 \includegraphics[width=0.5\columnwidth]{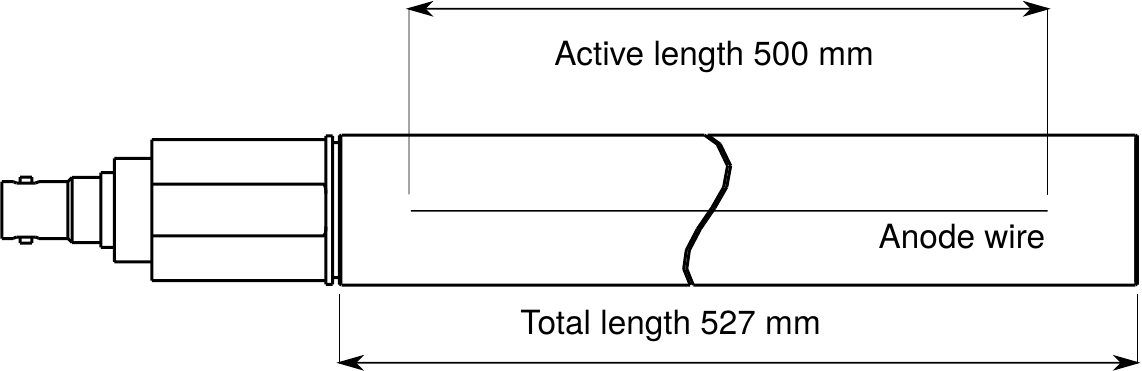}
\end{center}
\caption{Drawing of the \het\ tubes in ELIGANT-TN. The thickness of the anode wire is not to scale.\label{fig:tube_drawing}}
\end{figure}

The \ac{ELIGANT-TN} moderator for neutron counting contains 28 \het\ gas counters to detect the thermalised neutrons. After thermalization, the neutrons induce ${^{3}\mathrm{He}}(\mathrm{n}, \mathrm{p}){^{3}\mathrm{H}}$ reactions, with a $Q$ value of 765~keV, within the counting gas. As this is a two-body process, due to energy conservation, the kinetic energy of the fragments will be split inversely proportional to the mass, 3:1. This gives a proton with a kinetic energy of 573~keV and a triton with kinetic energy of 191~keV. The induced charge is then collected on the thin tensioned anode wire suspended in the centre of the gas counter tube. Each tube has a physical diameter of 25.4~mm and an active diameter of 24.4~mm. The total length of each tube is 527~mm, with an active length of 500~mm, shown in Figure~\ref{fig:tube_drawing}. The tubes are filled with \het\ gas at 12~atm pressure and a small amount of CO$_{2}$ 
as a quenching gas.

\begin{figure*}[ht!]
 \begin{center}
 \includegraphics[width=0.5\textwidth]{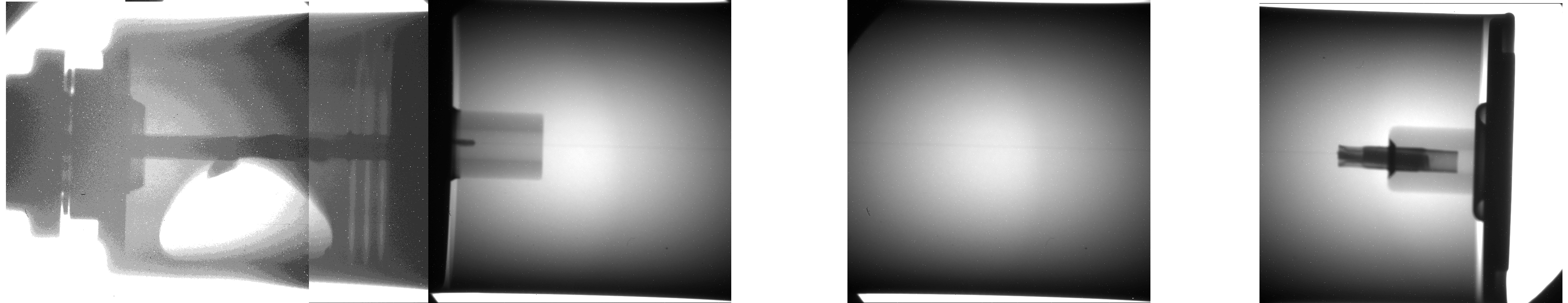}\\
 \includegraphics[width=0.7\textwidth]{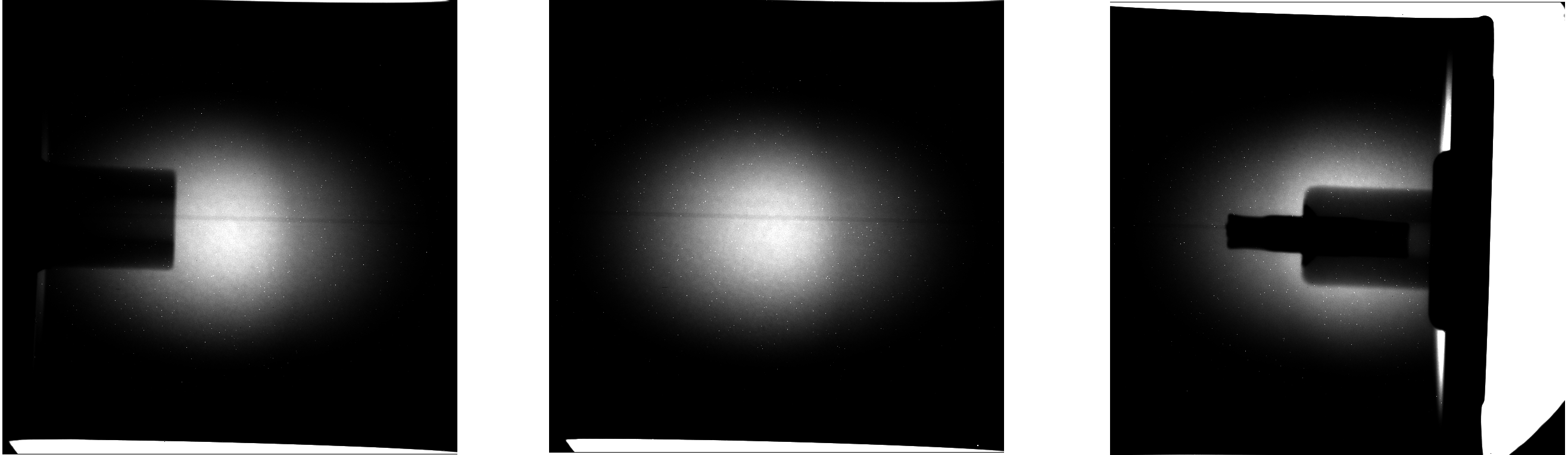}\\
\end{center}
\caption{X-ray of one of the \het\ tubes in ELIGANT-TN (top) with a high-contrast view on the anode wire (bottom).\label{fig:tube_xray}}
\end{figure*}

The internal structure of the tubes is shown in Figure~\ref{fig:tube_xray}, which was taken using the X-ray imaging laboratory at \ac{ELI-NP} \cite{Safca2022a,Stutman2023a}. This image was obtained with a distance between the source and the detector of 60~cm, a distance between the source and the tube of 45~cm, a source size of 7~$\mu$m with a voltage of 80~kV at 50~$\mu$A current. The exposure time for the image was 60~s. The technique of X-ray imaging was used to monitor the anode, as demonstrated in the lower part of Figure~\ref{fig:tube_xray}.

\section{High Voltage\label{sec:hv}}

The \ac{HV} is provided to the setup from an ISEG NHS 6030p \ac{NIM} unit. The ISEG NHS 6030p has six independent channels, where each channel's voltage and current control can be performed using the front panel directly on the \ac{NIM} unit or remotely via a \ac{USB} connection. This unit can provide a voltage up to 3~kV with a maximum current of $3$~mA, or equivalently, a maximum power of 9~W. The voltage is expected to be stable with a maximum ripple of 10~mV. The \ac{HV} is provided to the \het\ counters via \ac{SHV} connectors distributed via four pre-amplifier modules, see Section~\ref{sec:daq}.

\begin{figure}[ht!]
 \begin{center}
 \includegraphics[width=0.5\columnwidth]{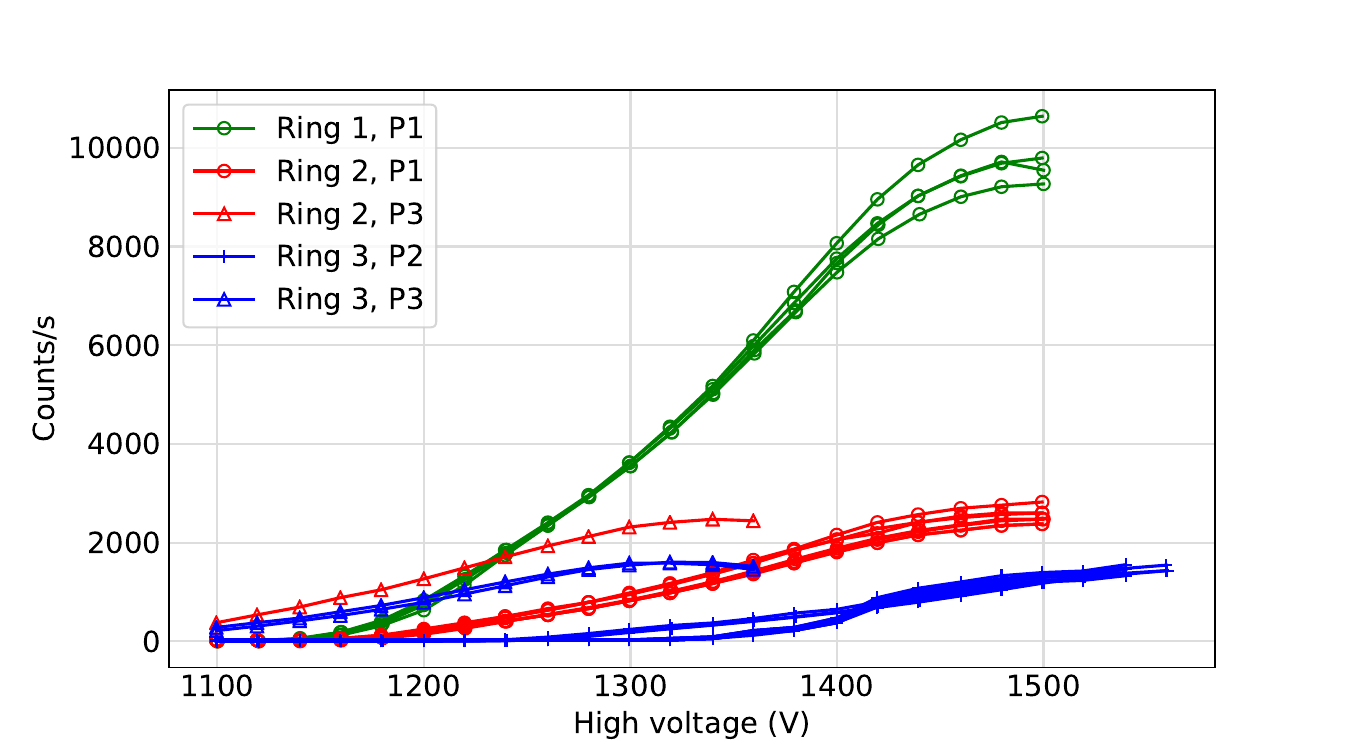}
\end{center}
\caption{Count rate as a function of high voltage of the \het\ counters distributed via three pre-amplifiers (circles, triangles, pluses), for the tubes in the inner ring (green), middle ring (red), and outer ring (blue).\label{fig:voltage_curves}}
\end{figure}

To determine the optimal \ac{HV} values for the \ac{ELIGANT-TN} counters, the count rate was evaluated as a function of \ac{HV} using a {\red{\ac{PuBe} composite}} source \cite{Soderstrom2021} with a neutron emission rate of $2.2\times10^{5}$ neutrons per second, shown in Figure~\ref{fig:voltage_curves}. For lower \ac{HV} values, the neutron count rate is strictly increasing with increasing \ac{HV}. However, the count rate increase will stop at a certain point, and the neutron count rate is independent of the \ac{HV}. To minimise systematic uncertainties in high-precision neutron cross-section measurements, we want to operate \ac{ELIGANT-TN} in the region that is insensitive to the \ac{HV} and, thus, the \ac{HV} from the ISEG NHS 6030p channels are set just above the ``knee'' in the \ac{HV} curve, according to standard procedures.

\section{Data Acquisition \label{sec:daq}}

The data from the \het\ tubes are extracted via the combined pre-amplifiers and \ac{HV} distribution units. The pre-amplifiers used for \ac{ELIGANT-TN} are the {\red{high-voltage versions}} of the original multichannel pre-amplifier MPR-16 by Mesytec. 
As the signal from \ac{ELIGANT-TN} are taken from the \ac{HV} output of the tubes, the large \ac{DC} voltage in the signal cable is removed by 
a capacitive coupling stage in the MPR-16. The bias voltage on the pre-amplifier input of each unit is filtered by an \ac{RC} filter with 10~M$\Omega$ resistance and 6.8~nF capacitance. 50~M$\Omega$ resistors then distribute the voltage to the 16 different channel inputs. The low-voltage for the pre-amplifiers is provided via a Mesytec four-channel MNV-4 \ac{NIM} power distribution module and provided via standard \ac{D-sub} connectors of type DE-9 both on the MNV-4 side and the MPR-16 side.

\begin{figure}[ht!]
 \begin{center}
 \includegraphics[width=0.5\columnwidth]{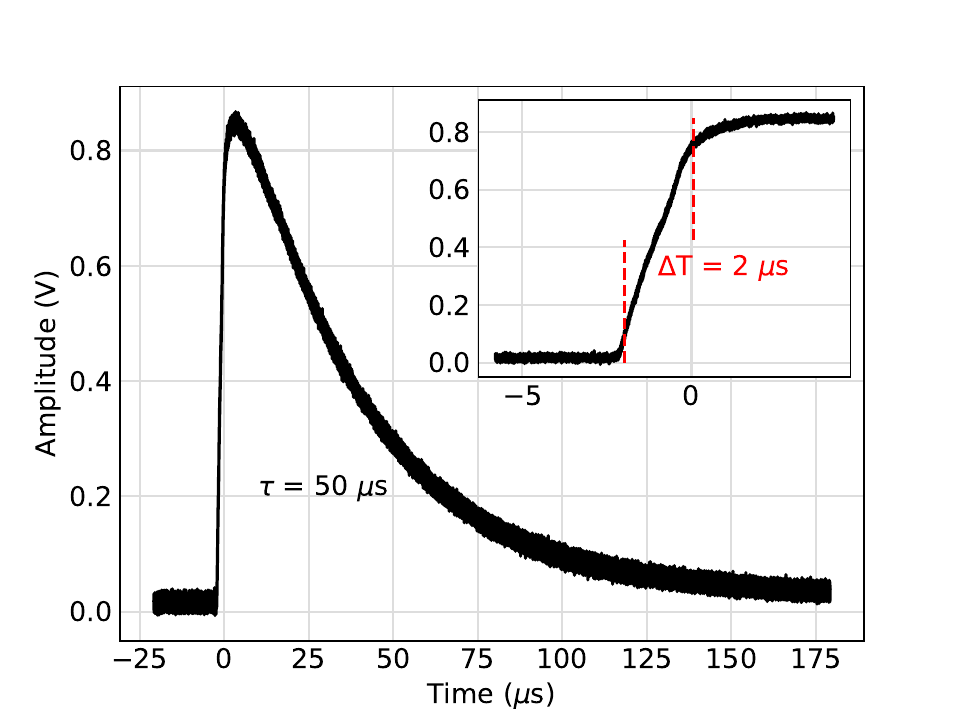}
\end{center}
\caption{Typical detector pulse charaacteristics after the pre-amplifiers used with ELIGANT-TN.\label{fig:one_tube_one_pulse}}
\end{figure}

\begin{figure}[ht!]
 \begin{center}
 \includegraphics[width=0.5\columnwidth]{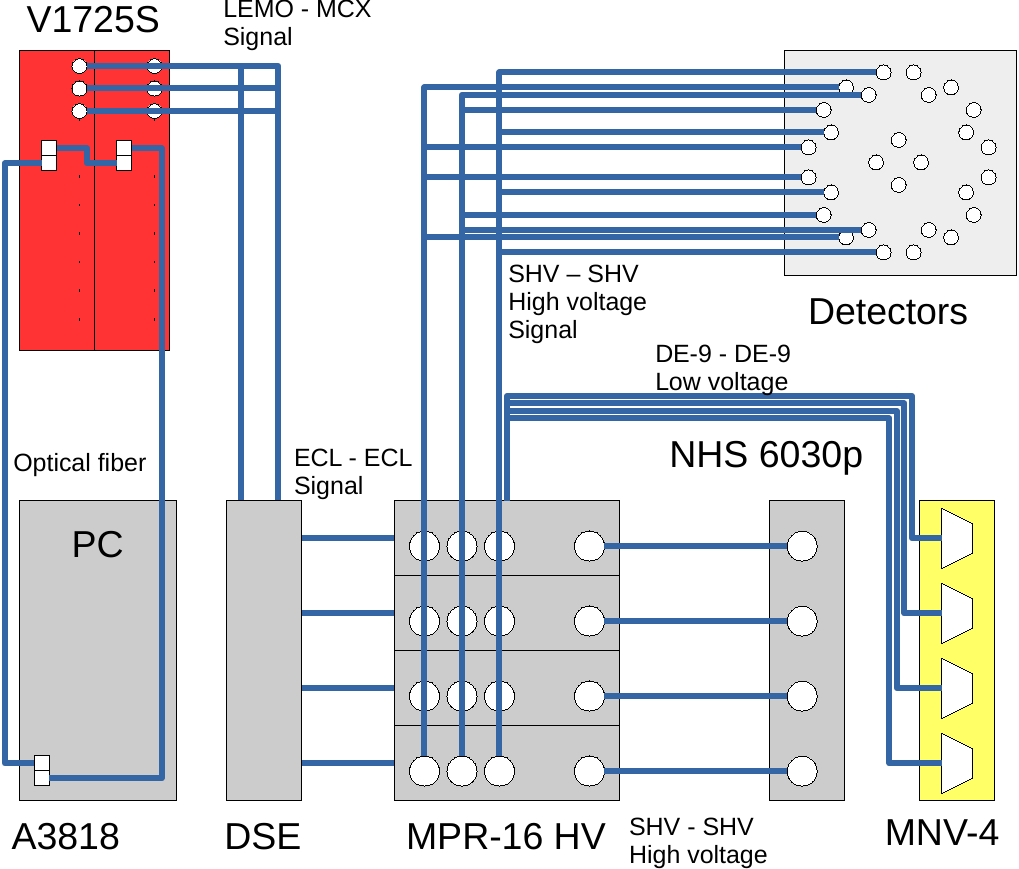}
\end{center}
\caption{Illustration of the readout and high-voltage electronic scheme for ELIGANT-TN. {\red{The detector signals pass via SHV cables to the MPR-16 HV pre-amplifiers, that are provided with high voltage from a NHS 6030p high precision versatile high voltage unit and powered by the MNV-4 power distribution and control module. The ECL signals from the pre-amplifier are fed to in-house differential-to-single ended converters (DSE) and the single-ended signals are taken via LEMO to MCX cables to the V1725S 125~Ms/s digitiser, connected via an optical link to the acquisition computer.}}\label{fig:electronic_setup}}
\end{figure}

The pre-amplifiers are carefully decoupled to protect against discharges that may occur within gas detectors. For \ac{ELIGANT-TN}, the differential output version of MPR-16 was chosen for better protection against ground noise. Some noise contribution in the pre-amplifiers will come from decoupling capacitors at high voltage. Thus, before any high-precision measurements are performed, the detectors and pre-amplifiers should be biased for a few hours. The pre-amplifiers have 
an input protection for positive bias to survive sparks even at 2000~V. 
This, however, has the adverse effect that both rise time and noise are increased, but it does not significantly affect the readout performance of \het\ tubes. The output of the pre-amplifiers is connected 
to two CAEN V1725 digitiser cards{\red{, one tube per input channel,}} with a sampling rate of 250~MS/s and a resolution of 14~bits via a set of {\red{in-house manufactured}} 
differential to single-ended converters. The need to include these boards in the chain is justified by the move towards a digital \ac{DAQ}, which requires unipolar pulses, compared to the fully analogue system that was in the initial design \cite{Camera2016} and could operate directly with the differential signals from the pre-amplifiers.
\begin{figure}[ht!]
 \begin{center}
 \includegraphics[width=0.5\textwidth]{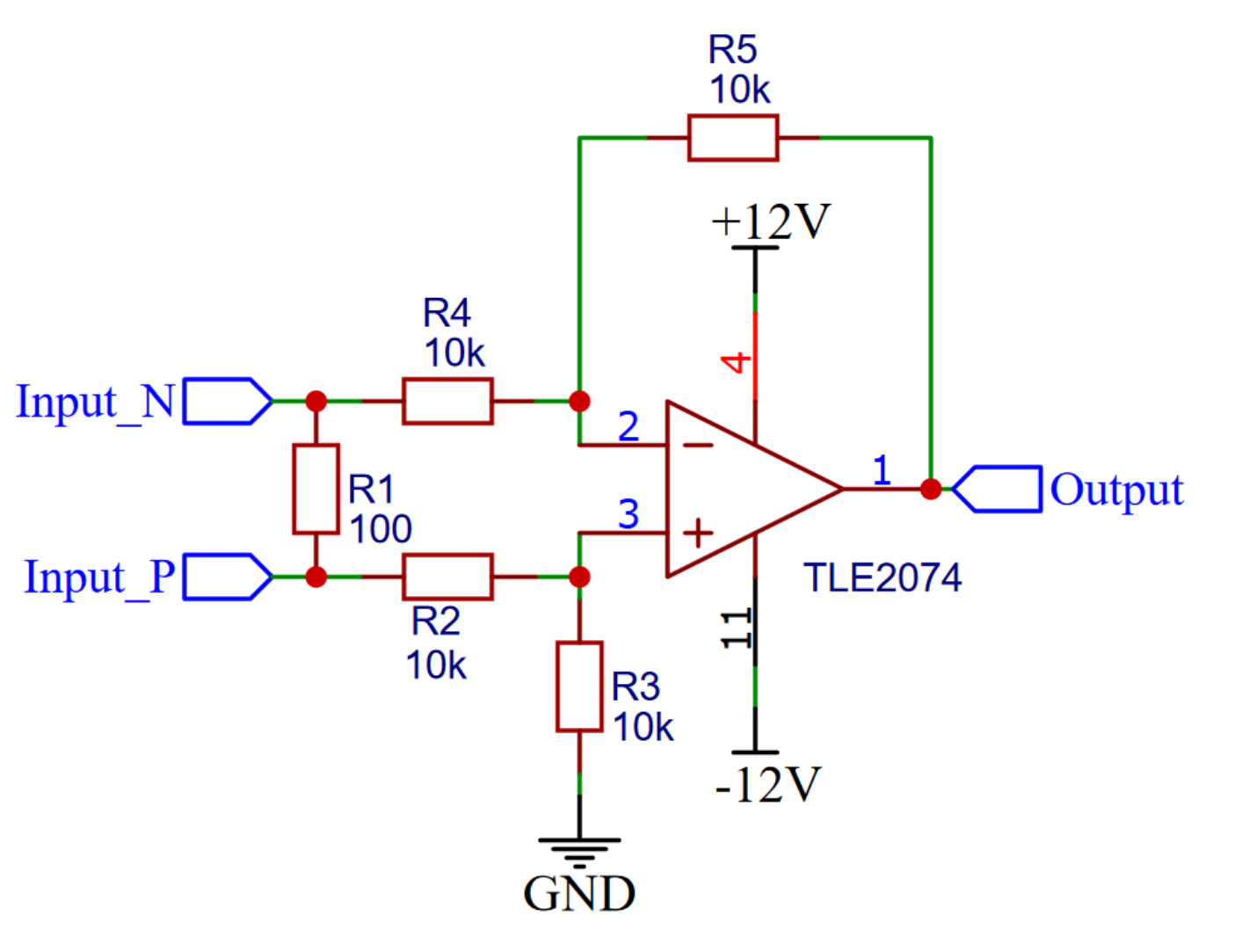}\\
 \includegraphics[width=0.5\textwidth]{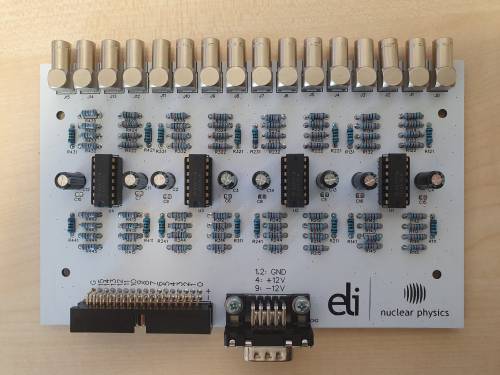}
\end{center}
\caption{{\red{(Top) Schematic drawing of the differential-single-ended converter. (Bottom) Photograph of one of the boards produced.\label{fig:converters}}}}
\end{figure} 
{\red{These converter boards are centred around commercially available operational amplifiers arranged in a configuration to accept differential signals and provide a bipolar signal referenced to the ground. Each board has 16 channels and accepts the differential signals via a ribbon cable, using the same 34-pin socket as the pre-amplifier. Outputs are presented on LEMO 00 sockets, and power is provided through the same DE-9 connector to ease integration with existing laboratory equipment.
A single-channel schematic is presented in Figure~\ref{fig:converters}. The differential input is terminated with resistor R1. Resistors R2 to R5 form a network that presents the following voltages at the operational amplifier inputs:
\begin{align}
  V_{\mathrm{in}}+ &= \mathrm{Input\_P}/2\\
  V_{\mathrm{in}}- &= (\mathrm{Output} + \mathrm{Input\_N}) / 2
\end{align}
The feedback in the operational amplifier drives the output such that $V_{\mathrm{in}}+ = V_{\mathrm{in}}-$, thus making $\mathrm{Output} = \mathrm{Input\_P} - \mathrm{Input\_N}$. Output signals from the pre-amplifiers have a fast rise time; see Figure~\ref{fig:one_tube_one_pulse}; thus, the operational amplifier slew rate is important in preserving signal integrity for processing in the digitiser. The Texas Instruments TLE2074 has a slew rate of 45V/$\mu$s and includes four operational amplifiers in a 14-pin package, simplifying the \ac{PCB} layout. A picture of the completed converter is presented in Figure~\ref{fig:converters}.}}
A drawing of the complete electronic layout is shown in Figure~\ref{fig:electronic_setup}, indicating how the different components interconnect with each other.

\begin{figure}[ht!]
 \begin{center}
 \includegraphics[width=0.5\textwidth]{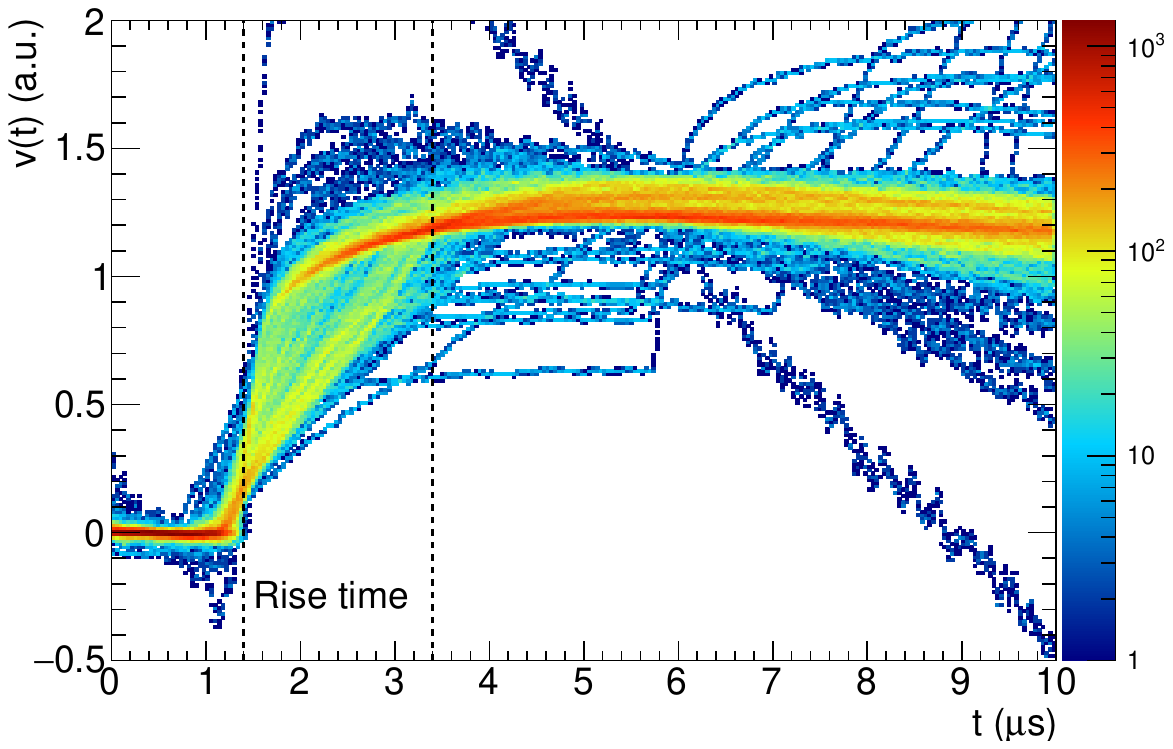}\\
 \includegraphics[width=0.5\textwidth]{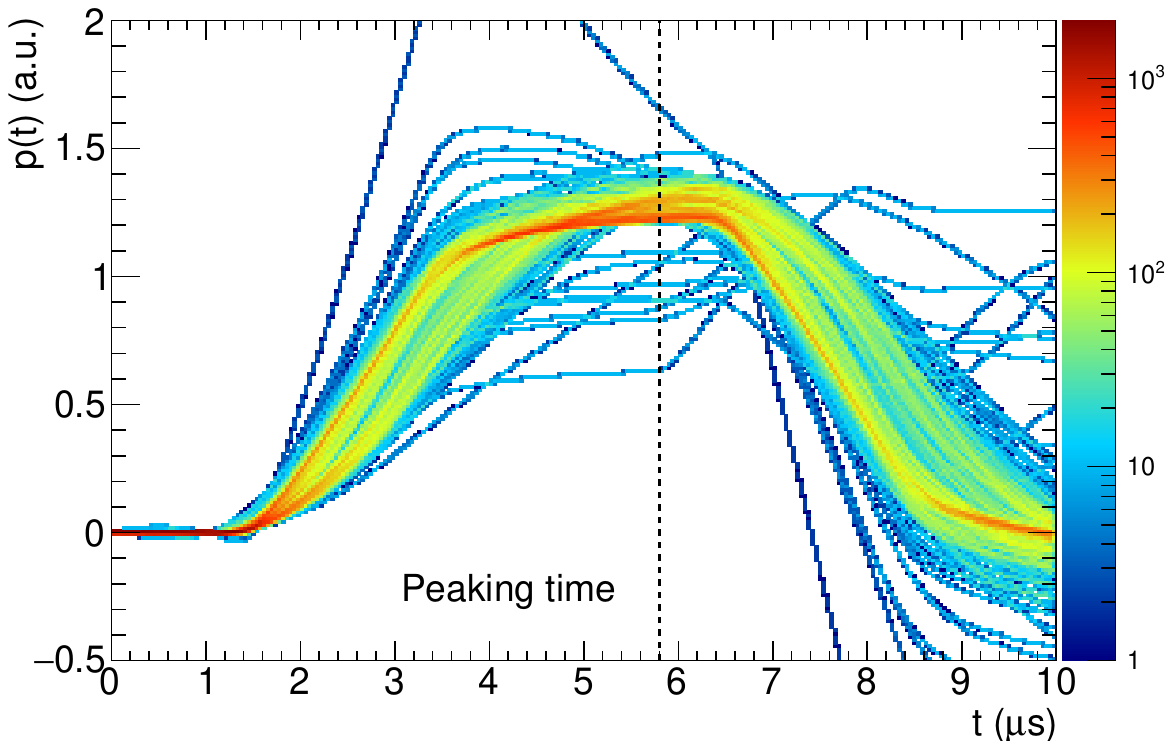}
\end{center}
\caption{Digitised waveforms from one of the ELIGANT-TN \het\ tubes via a simple 34 pin flat header connector to 16 Lemo connectors cable converter at the pre-amplifier output (top) and final trapezoidal pulses for energy extraction (bottom).\label{fig:cube_pulse}}
\end{figure}

The signal processing in the V1725 digitisers is performed using the \ac{DPP-PHA} firmware, commonly employed for pre-amplifier signals with a long decay time, for example, from \ac{HPGe} detectors. This firmware includes a trapezoidal filtering algorithm to extract the energy information in the signal, the specifics of which will determine count rate capabilities, energy resolution, and similar. We evaluated these parameters for a data set of waveforms, $v(n)$, collected using a {\red{\ac{PuBe}}} source \cite{Soderstrom2021}, where $n$ is a sample point corresponding to a time $t=4n$~ns. For the evaluation, a simplified version of the algorithm proposed in References~\cite{Jordanov1994a,Jordanov1994b} was used, where the filtered waveform $p(n)$ was obtained from
\begin{align}
    d^{k,l}(n)&=v(n)-v(n-k)-v(n-l)+v(n-k-l),\\
    p(n)&=p(n-1)+d^{k,l}(n).
\end{align}
In this case, the parameters $k$ and $l=k+m$ correspond to the rise time of the trapezoidal pulse ($k$) and the length of the flat top of the trapezoid ($m$). Our algorithm simplifies by excluding pole-zero correction in the pulse treatment due to the slightly inconsistent pulse shapes, shown in Figure~\ref{fig:cube_pulse}.
10~000 waveforms were selected for the pulse-shape analysis and normalised to the same integral value for visualisation. In addition to the clean waveforms, this data set also contained examples of pile-up pulses, pulses with distorted rise time, and pulses with distorted decay time. A trapezoidal rise-time value of $k=2$~$\mu$s was selected for a time range where most pulses have 10\%-90\% rise time within the shaping-time constant. Due to the various shapes in the leading edge of the signal, a time constant of $m=3$~$\mu$s was chosen for the flat top of the trapezoid, see Figure~\ref{fig:cube_pulse}. With these time constants, each \het\ tube should be able to handle a time between pulses of approximately $10$~$\mu$s without suffering from pile-up, and the counting rate in a full-scale experiment should be adjusted accordingly. {\red{The \ac{DPP-PHA} firmware in the V1725 digitizers can be set to identify these pile-up events by the triggering pattern within a given time-window to flag them as pile-up candidates. When a pile-up event is detected, the firmware can, depending on the time widow between the triggers, reconstruct both energies, reconstruct only the first energy, or reject the event completely. If the time difference between the triggers is below the sensitivity limit, the event will be recorded as a single event with the sum energy. As the typical count rates in the inner ring, Ring~A, contains approximately 60\% of the events, or 15\% per ring, the total array is expected to be able to handle several hundreds of kHz count rate in one-neutron experiments with negligible pile-up. For neutron-multiplicity measurements, however, the count rate will be limited by the moderation time with significantly stricter, by several orders of magnitude, limitations. This possibility to handle very high count rates is one of the major advantages with a digital system, as shown also in other experiments \cite{Aogaki2022,Aogaki2023,AogakiUnp}.}}

The data is collected using the in-house developed \ac{DAQ} system \ac{DELILA}, which has been discussed in detail in References~\cite{Aogaki2022,Aogaki2023,AogakiUnp}.

\section{{\sc{Geant4}} and MCNP Simulations}

The \ac{ELIGANT-TN} setup was simulated using a \geant\ \cite{Agostinelli2003} implementation based on the \ac{BRIKEN} code \cite{TarifenoSaldivia2017,Phong2020,BRIKENG4} as, due to the complexity of the \ac{BRIKEN} setup \cite{TarifenoSaldivia2017,Tain2018,TolosaDelgado2019,Phong2020} consisting of \ac{HPGe} detectors, silicon detectors, and several different types of \het\ tubes from RIKEN, \ac{UPC}, and \ac{ORNL}, is well developed and already tested both with respect to experimental data and \ac{MCNP} simulations. In this work, \geant\ version 4.10.04.p01 was used, as \geant\ versions lower than 4.10.0 contain a bug in the interpolation routines for thermal neutrons \cite{TarifenoSaldivia2017}.

\begin{figure}[ht!]
 \begin{center}
 \includegraphics[width=0.5\columnwidth]{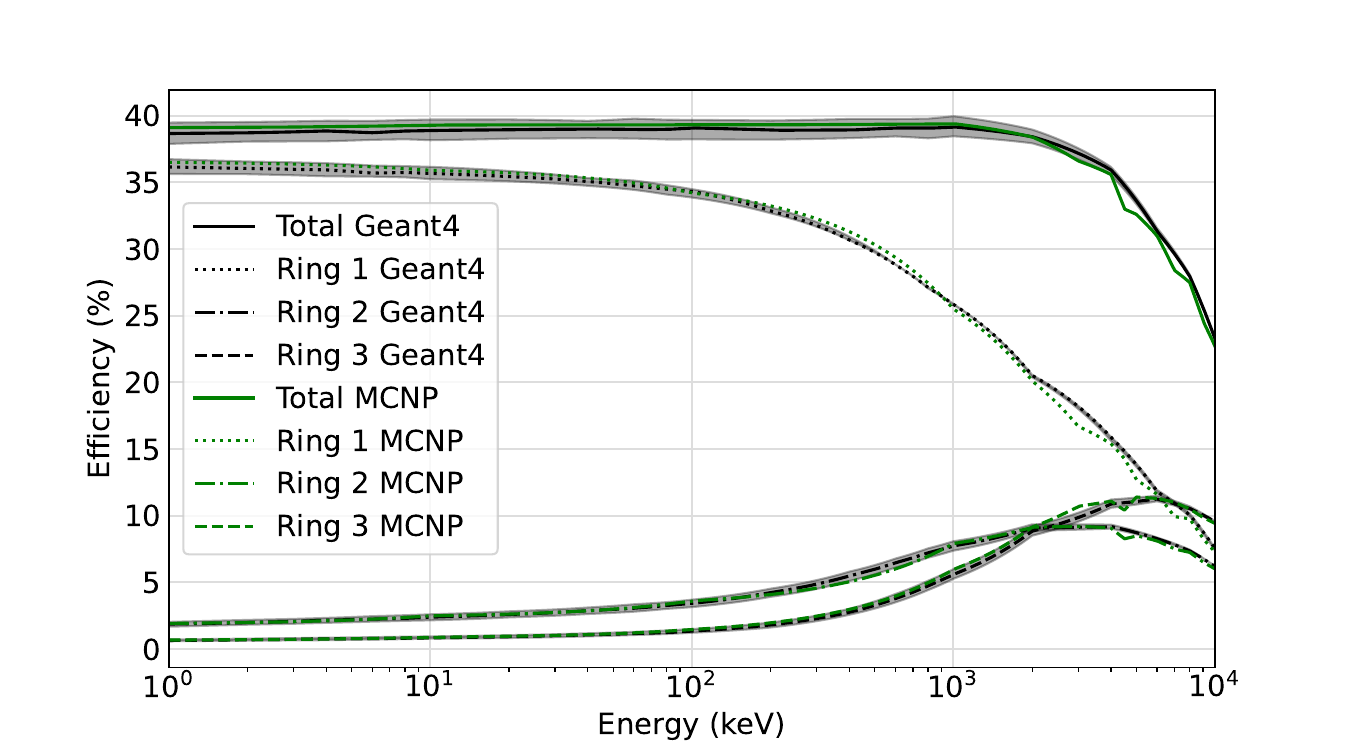}\\
 \includegraphics[width=0.5\columnwidth]{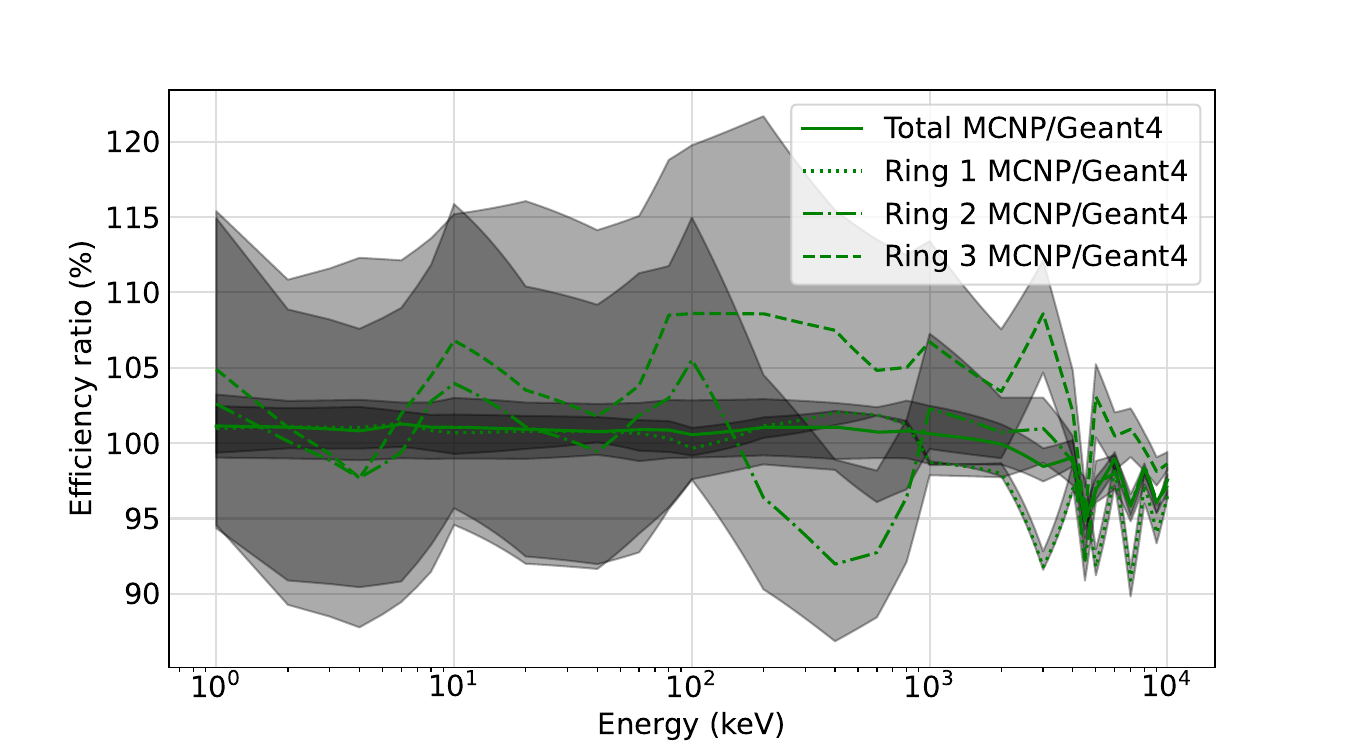}
\end{center}
\caption{Simulated efficiency of the ELIGANT-TN setup for isotropic, monoenergetic neutrons obtained from the \geant\ and MCNP codes. {\red{The black shaded areas correspond to a variation of 2\% in moderator density.}}\label{fig:simulated_efficiencies}}
\end{figure}

The \ac{HDPE} matrix was implemented with the material \texttt{TS\_H\_of\_Polyethylene} with a density of 0.95~g/cm$^{3}$, as defined in the \texttt{G4NeutronHPThermalScatteringNames} class of the high-precision thermal neutron scattering physics libraries, \texttt{NeutronHP}, of \geant. These libraries not only take the nuclear processes into account when transporting close to thermal neutrons but also include energy loss or gain from interactions, including translational, rotational, and vibrational modes in the molecular structure of the material. Thus, the temperatures and states of the material were also defined as solid for the \ac{HDPE} and gas for the \het. {\red{The main parameter in the detector construction that the efficiency has a significant sensitivity to is the density of the \ac{HDPE}. Typical \ac{HDPE} densities varies between 0.93~g/cm$^{3}$ to 0.97~g/cm$^{3}$, and we show the impact on possible density variations by performing the simulations with the nominal density, and densities $\pm 2$\%.}} The tubes themselves were defined as consisting of stainless steel with a density of 8.03~g/cm$^{3}$ and a composition of 66.495\% iron, 20\% chromium, 10.5\% nickel, 2\% manganese, 0.75\% silicon, 0.1\% nitrogen, 0.08\% carbon, 0.045\% phosphorus, and 0.03\% sulphur. A total length of 55~cm and a diameter of 2.54~cm of solid stainless steel were used for each tube, neglecting mechanical details in the connectors. The inner material was defined as gaseous \het\ at a temperature of 298.15~K and a pressure of 12~bar. 
For the \het\ gas, an active length of 50~cm and an active diameter of 2.438~cm were used. Implementing the \het\ tubes did not account for the efficiency loss from edge effects, as discussed recently in Reference~\cite{Peplowski2020}. Another simplification in the \geant\ simulation was the complete enclosure of the \het\ tubes in the \ac{HDPE} moderator. Thus, the entry holes at one edge of the \ac{HDPE} matrix were not explicitly included. The neutron efficiency curve of the total \ac{ELIGANT-TN} setup, as well as the individual rings, was obtained by generating 1~000~000 isotropic monoenergetic neutrons in energy steps ranging from 0.1~keV to 10~MeV and is shown in Figure~\ref{fig:simulated_efficiencies}.

\begin{figure}[ht!]
 \begin{center}
 \includegraphics[width=0.5\columnwidth]{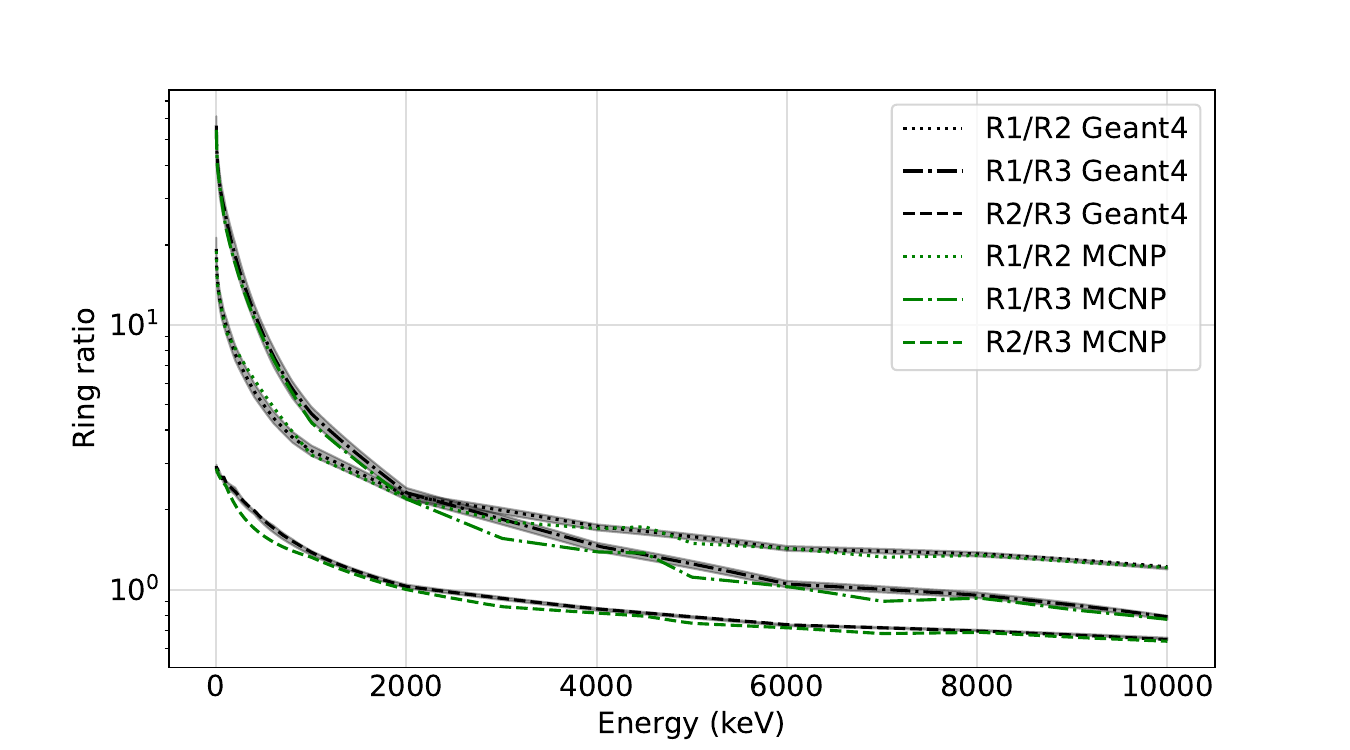}
\end{center}
\caption{Simulated energy dependence of the ring ratio for the ELIGANT-TN setup with isotropic, monoenergetic neutrons, as obtained from the \geant\ and MCNP codes. {\red{The black shaded areas correspond to a variation of 2\% in moderator density.}}}\label{fig:simulated_ring_ratio}
\end{figure}

While \ac{ELIGANT-TN} aims to be a neutron counter for cross-section measurements, information about the average energy of the neutrons emitted can be obtained from the so-called ring-ratio technique \cite{Berman1975}. The central concept of this approach is to use the fact that different amounts of moderator shielding in front of the \het\ counter will affect the efficiency differently for different neutron energies. Thus, the ratios of the number of counts in the different rings should, in the first order, correspond to the mean energy of the emitted neutrons. For monoenergetic neutrons, such a relation is shown in Figure~\ref{fig:simulated_ring_ratio} for the simulated data.

\begin{figure}[ht!]
 \begin{center}
 \includegraphics[width=0.5\columnwidth]{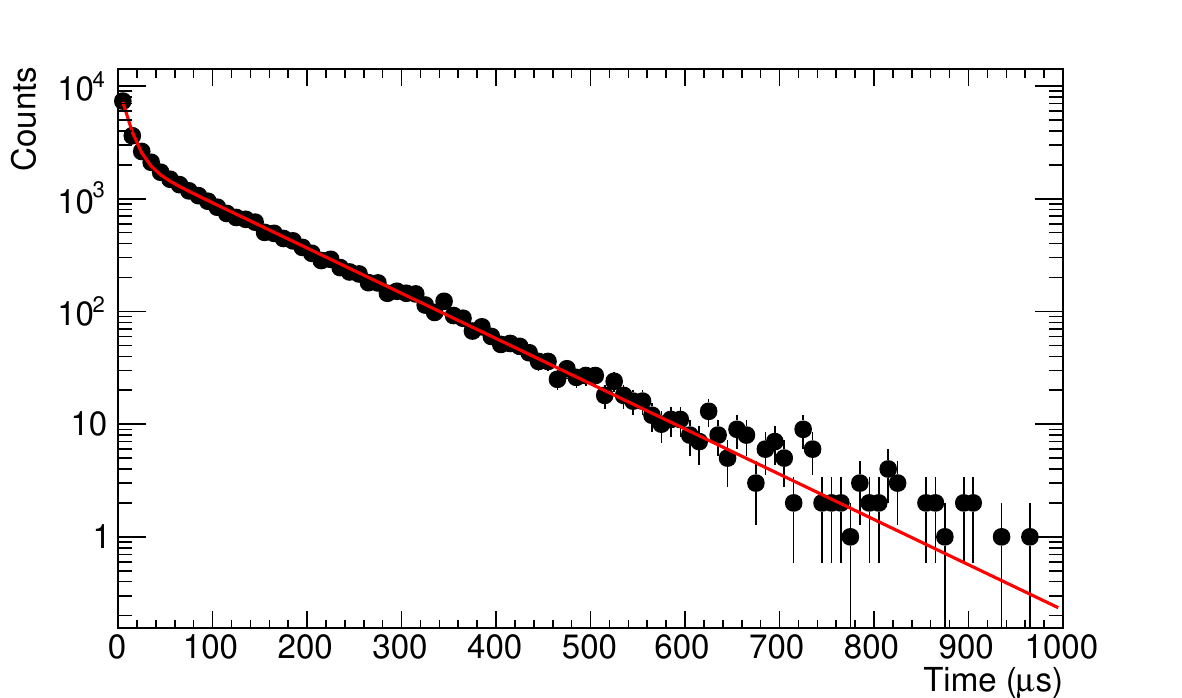}
\end{center}
\caption{\geant\ simulated time spectrum of the detected neutron from a PuBe neutron source. Two exponential decay components have been fitted to the simulated data, with time constants {\red{$\tau_{1}=10.27(35)$~$\mu$s}} and {\red{$\tau_{2}=108.2(9)$~$\mu$s}}, corresponding to approximately {\red{25.0(5)}}\% of the integrated counts in component 1 and approximately {\red{75.0(6)}}\% of the integrated counts in component 2.
\label{fig:pube_time_spectrum}}
\end{figure}

In order to define events for data analysis which contain as large an amount as possible of actual correlated neutrons while, at the same time, reducing the background as much as possible, the time distribution of the detected neutrons following moderation has been studied using the \geant\ simulations. A typical simulated time distribution from a {\red{\ac{PuBe}}} neutron source is shown in Figure~\ref{fig:pube_time_spectrum}.
This time spectrum contains two characteristic exponential features, and to describe these features, we have applied the function
\begin{equation}
    f(t) = a_{1}\mathrm{e}^{-t/\tau_{1}}+a_{2}\mathrm{e}^{-t/\tau_{2}} \label{eq:timedist},
\end{equation}
where $a_{1}$ is the value at $t=0$ for component 1 and $a_{2}$ is the value at $t=0$ for component 2, with time constants $\tau_{1}$ and $\tau_{2}$, respectively. These time components will determine the maximum count rate possible for the setup with the \ac{ELI-GBS} beams. For practical purposes in the context of \ac{ELIGANT-TN}, the beam can be considered continuous in situations where correlated neutron detection experiments are performed. This could, for example, be when we are above the two-neutron separation threshold and want to compare one-neutron and two-neutron emission cross-sections with the direct neutron-multiplicity sorting technique \cite{Utsunomiya2017}, or for photofission experiments. Assuming a pure Poisson distribution, we can calculate the probability for more than one neutron-emitting reaction to happen within the time window, as well as the number of missed events due to them being outside the time window, from Equation~(\ref{eq:timedist}). A selected number of cases, listed for illustrative purposes, are presented in Table~\ref{tab:poisson_properties}.
\begin{table}[ht]
\caption{Estimated correction factors from Poisson statistics for ELIGANT-TN experiments with the ELI-GBS beam. The reaction rate denotes the number of reactions in the target, $\Delta T$ the time window for considering neutrons in one event correlated, $P_{n>1}$ the probability for more than one neutron-emitting reaction to happen within the time window, and Loss denotes the average number of missed events due to them being outside the time window.\label{tab:poisson_properties}}
 \begin{tabular*}{\columnwidth}{@{\extracolsep{\fill}}cccc}
\hline	
Reaction rate	&	$\Delta T$	&	$P_{n>1}$ &	Loss\\	
 (Hz)	&	($\mu$s)	& (\%)	& (\%)  \\	
\hline
10 & 300 & 0.30 & 5.63\\
100 & 300 & 2.96 & 5.63\\
1000 & 300 & 25.92 & 5.63\\
10000 & 300 & 95.02 & 5.63\\
10 & 500 & 0.50 & 0.89\\
100 & 500 & 4.88 & 0.89\\
1000 & 500 & 39.35 & 0.89\\
10000 & 500 & 99.33 & 0.89\\
10 & 1000 & 1.00 & 0.01\\
100 & 1000 & 9.52 & 0.01\\
1000 & 1000 & 63.21 & 0.01\\
10000 & 1000 & 100.00 & 0.01\\
\hline																			
 \end{tabular*} 
\end{table}

The \ac{MCNP} code MCNPX (software package C00810 MNYCP 01) \cite{mcnp} was used to simulate the neutron transport. The geometrical simulation model was based on the same dimensions for the \ac{HDPE} shielding block and \het\ tube parameters as presented in Section~\ref{sec:mechanicaldesign}. Material densities were 0.95 g/cm$^{3}$ for \ac{HDPE} and 8.03 g/cm$^{3}$ for stainless steel. The active region of each \het\ tube was filled with \het\ gas with a density of $2.479 \times 10^{-4}$ atoms/(b-cm). The neutron detection efficiency was obtained by generating $10^{7}$ isotropic monoenergetic neutrons in energy steps ranging from 1~keV to 10~MeV using the source definition card. The \ac{MCNP} \ac{F4} tally calculated the average neutron fluence per simulated source-neutron over the active region of each \het\ tube. The efficiency is acquired directly for each tube from the \ac{F4} tallies with the FM multiplier card, which takes into account the number of (n,p) reactions in \het. The results from these simulations are shown together with the \geant\ results in Figures~\ref{fig:simulated_efficiencies} and \ref{fig:simulated_ring_ratio}. While the overall agreement between the two approaches is good, the difference is explicitly shown in Figure~\ref{fig:simulated_efficiencies}. Here, it can be seen that while the total efficiencies agree within 3\% over the full energy range, the main difference between the two simulations is in the efficiencies of the individual rings, where the difference for Ring~{\red{B}} and Ring~{\red{C}} in the worst cases is as large as 8\%. 

\section{{\red{Source tests}}}

To characterise the detectors' performance with data and verify consistency between data and simulations, a set of source and in-beam measurements has been carried out. The tubes were mounted for these measurements as shown in Table~\ref{tab:det_table}.
\begin{table}[ht]
\caption{Configuration of the ELIGANT-TN \het\ counters in the source and in-beam measurements reported here. The detector ID and serial numbers are listed, together with the pre-amplifier used and the high-voltage (HV) applied as determined from the HV optimisation in Section~\ref{sec:hv}. Also, the specified high voltage from the manufacturer, HV$_{\mathrm{spec}}$, is listed for reference, as well as the precise angles and distances of each tube from the centre of the target.\label{tab:det_table}}
\scriptsize
 \begin{tabular*}{\columnwidth}{@{\extracolsep{\fill}}ccccccc}
\hline																			
ID	&	Ser. \#	&	PA &	HV & HV$_{\mathrm{spec}}$ & $\theta$ & $r$\\
   &  &  & (V) & (V) & & \\	
\hline																A1 & 1973 &	P1 & 1500 & 1400 & 90 & 5.9 \\	
A2 & 1974 &	P1 & 1500 & 1400 & 0 & 5.9 \\	
A3 & 1975 &	P1 & 1500 & 1400 & -90 & 5.9 \\	
A4 & 1976 &	P1 & 1500 & 1400 & 180 & 5.9 \\	
B1 & 1977 &	P1 & 1500 & 1400 & 67.5 & 13.0 \\	
B2 & 1978 &	P1 & 1500 & 1400 & 22.5 & 13.0 \\	
B3 & 1979 &	P1 & 1500 & 1400 & -22.5 & 13.0 \\	
B4 & 1980 &	P1 & 1500 & 1400 & -67.5 & 13.0 \\	
B5 & 1981 &	P1 & 1500 & 1400 & -112.5 & 13.0 \\	
B6 & 1982 &	P1 & 1500 & 1400 & -157.5 & 13.0 \\	
B7 & 1983 &	P3 & 1350 & 1400 & 157.5 & 13.0 \\	
B8 & 1984 &	P1 & 1500 & 1400 & 112.5 & 13.0 \\	
C1 & 1985 &	P2 & 1540 & 1400 & 78.8 & 15.5 \\	
C2 & 1986 &	P2 & 1540 & 1400 & 56.3 & 15.5 \\	
C3 & 1987 &	P2 & 1540 & 1400 & 33.7 & 15.5 \\	
C4 & 0241 &	P2 & 1540 & -  & 11.2 & 15.5 \\	
C5 & 1989 &	P2 & 1540 & 1400 & -11.2 & 15.5 \\	
C6 & 1990 &	P2 & 1540 & 1400 & -33.7 & 15.5 \\	
C7 & 1992 &	P2 & 1540 & 1400 & -56.3 & 15.5 \\	
C8 & 1993 &	P2 & 1540 & 1400 & -78.8 & 15.5 \\	
C9 & 2000 &	P2 & 1540 & 1400 & -101.2& 15.5 \\	
C10 & 2001 & P2 & 1540 & 1400 & -123.7 & 15.5 \\	
C11 & 2002 & P2 & 1540 & 1400 & -146.3 & 15.5 \\	
C12 & 1997 & P3 & 1350 & 1260 & -168.8 & 15.5 \\	
C13 & 1998 & P3 & 1350 & 1260 & 168.8 & 15.5 \\	
C14 & 1994 & P2 & 1540 & 1500 & 146.3 & 15.5 \\	
C15 & 1995 & P2 & 1540 & 1500 & 123.7 & 15.5 \\	
C16 & 1996 & P2 & 1540 & 1500 & 101.2 & 15.5 \\	
\hline
\end{tabular*} 
\end{table}


A first set of source measurements was performed using a {\red{\ac{PuBe}}} source available at the institute 
\cite{Soderstrom2021}. The listed activity of the {\red{\ac{PuBe}}} source was $2.27\times10^{5}$~n/s, according to the import certificate to Romania dated May 7, 1976. 
The active volume was in the form of a cylinder with an 8~mm diameter and 8~mm height.

The \ac{ELIGANT-TN} setup was read out by two CAEN V1725 digitisers with PHA firmware, revision 4.15\_139.08. The data was collected for 10~minutes, giving $4.90\times10^{7}$ counts in the full array. In these measurements, the data was collected with a \ac{DAQ} computer running the proprietary CoMPASS software by CAEN via an optical link through a \ac{PCIe} board. The distribution of these counts within the rings was $2.35\times10^{7}$ counts in Ring~{\red{A}}, $1.20\times10^{7}$ counts in Ring~{\red{B}}, and $1.35\times10^{7}$ counts in Ring~{\red{C}}. {\red{The typical rejection rates due to pileup was 1.42\% for Ring~A, 0.68\% for Ring~B, and 0.43\% for Ring~C. This was the main source of dead-time in the measurement.}}

\begin{figure}[ht!]
 \begin{center}
 \includegraphics[width=0.5\columnwidth]{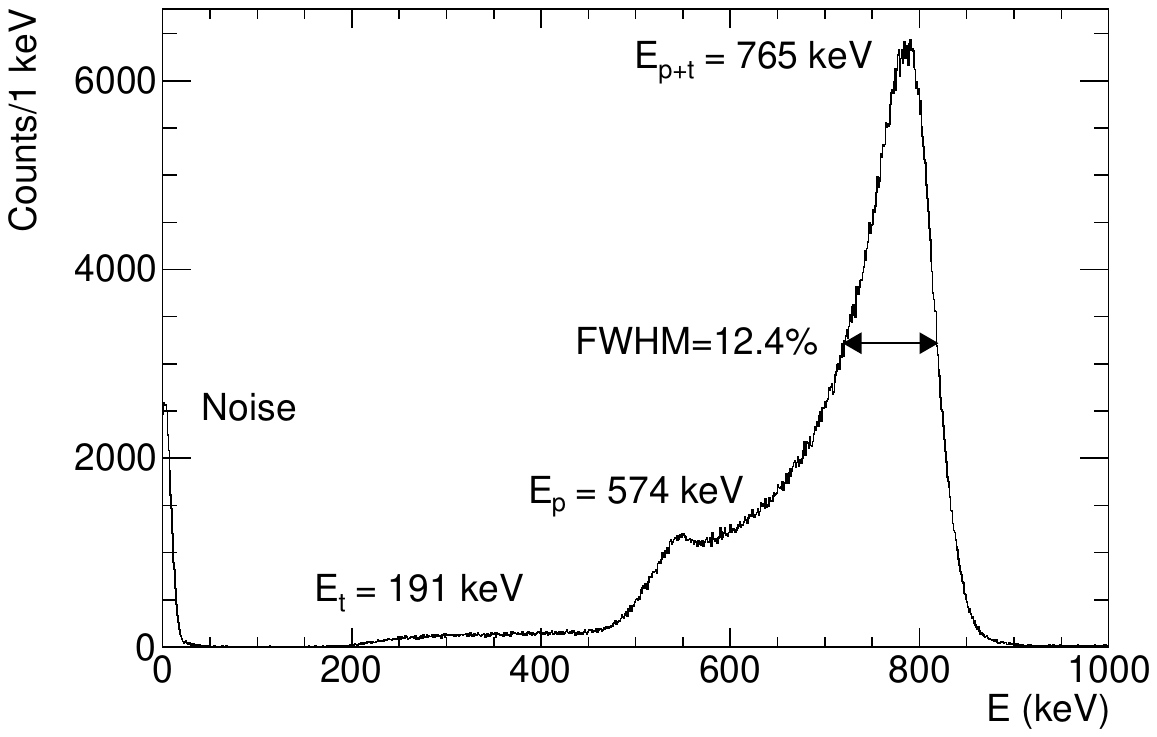}
\end{center}
\caption{Typical pulse-height spectrum from one of the \het\ counters using a PuBe source and the CoMPASS software.\label{fig:he3_spectrum}}
\end{figure}

A typical pulse-height spectrum collected with the {\red{\ac{PuBe}}} source is shown in Figure~\ref{fig:he3_spectrum}, where the characteristic features of a ${}^{3}$He pulse-height spectrum are visible.
At an energy of 765~keV, we have the peak corresponding to the total energy collection, with a width of 12.4\% \ac{FWHM}. In addition, the characteristic wall-effect thresholds, when the proton and triton daughter products of the capture reaction, with kinetic energies of 573~keV and 191~keV, respectively, collide with the wall of the counter. At the lowest energies, we can see the typical noise from $\gamma$ rays that partially deposit a small amount of energy in the ${}^{3}$He gas.

\begin{figure}[ht!]
 \begin{center}
 \includegraphics[width=0.5\columnwidth]{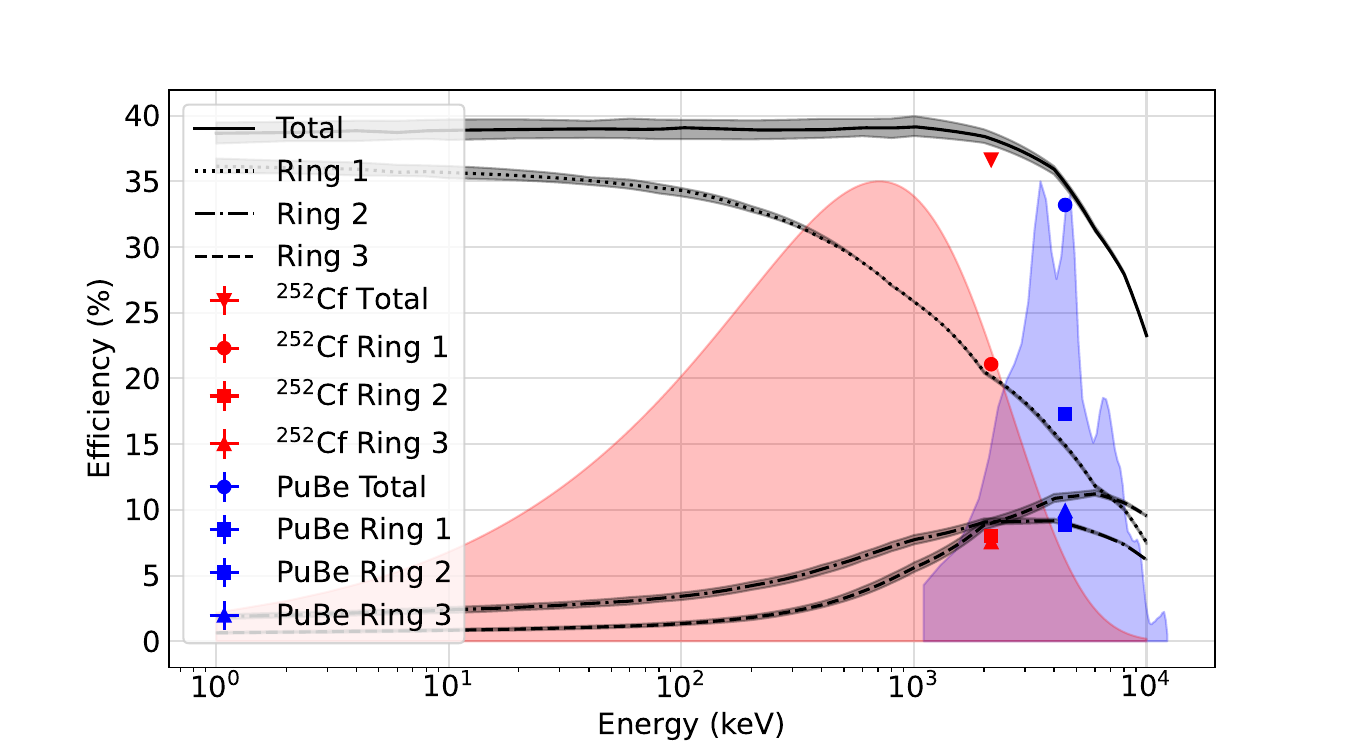}
\end{center}
\caption{\geant\ simulated efficiency of the ELIGANT-TN setup together with source measurements using a ${}^{252}$Cf source and a PuBe source. {\red{The ${}^{252}$Cf and PuBe energy data points correspond to the mean value of the distribution, and the complete energy distribution for the respective sources are shown as shaded regions. The black shaded areas correspond to a variation of 2\% in moderator density.}} \label{fig:efficiency_wpube}}
\end{figure}

Using the evaluated intensity and energy spectrum from the {\red{\ac{PuBe}}} source {\red{\cite{Soderstrom2021}}}, we can evaluate the efficiency of \ac{ELIGANT-TN} within this energy range and compare this to the simulated total efficiency and the efficiencies of the individual rings. The results of this evaluation are shown in Figure~\ref{fig:efficiency_wpube}.
The overall agreement between measurements and simulations is good, with the broad energy distribution being the primary source of uncertainty, especially since the {\red{\ac{PuBe}}} neutron-energy spectrum is well outside of the flat-efficiency region.

\begin{figure}[ht!]
 \begin{center}
 \includegraphics[width=0.5\columnwidth]{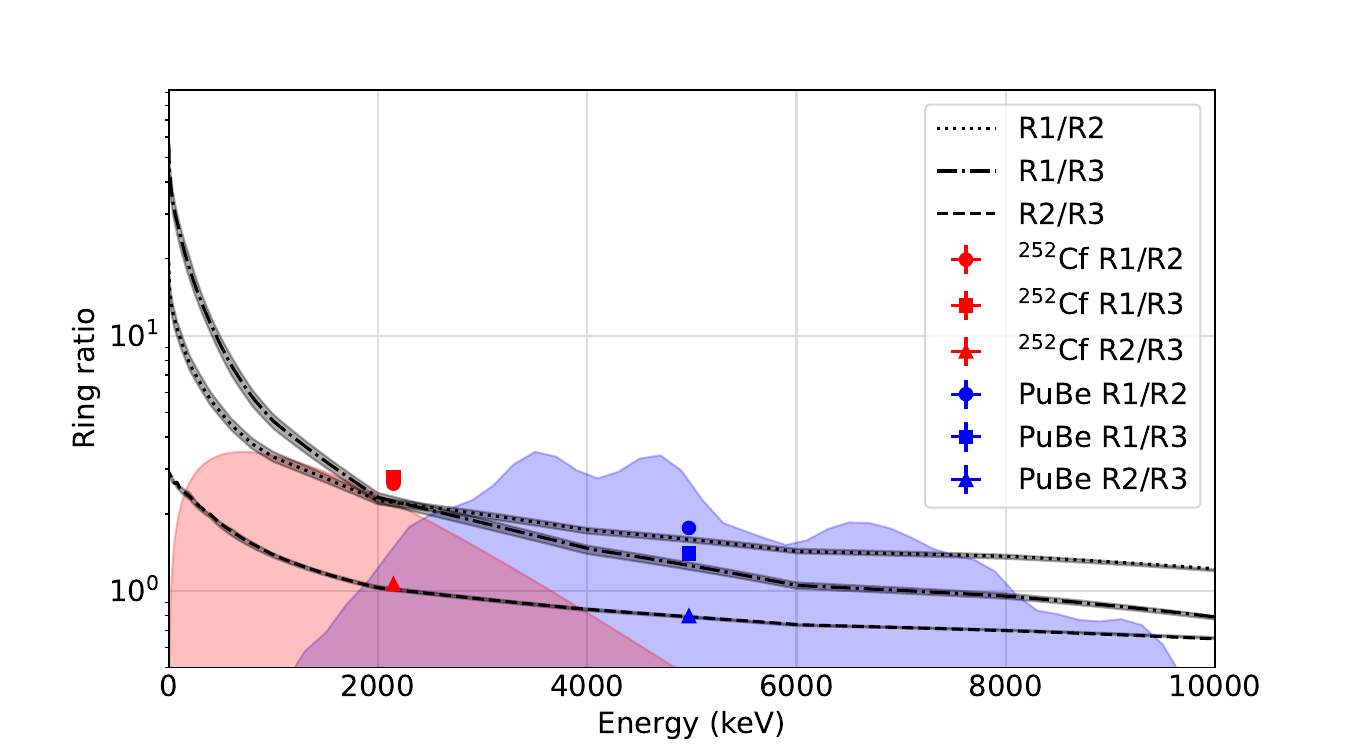}
\end{center}
\caption{Measured ring ratios from a ${}^{252}$Cf neutron source and a PuBe neutron source compared with monoenergetic \geant\ simulations. The energy uncertainties correspond to the approximate energy distributions from the sources. {\red{The ${}^{252}$Cf and PuBe energy data points correspond to the mean value of the distribution, and the complete energy distribution for the respective sources are shown as shaded regions. The black shaded areas correspond to a variation of 2\% in moderator density.}}\label{fig:ring_ratio_wpube}}
\end{figure}

 While \ac{ELIGANT-TN} works with neutrons moderated to thermal energies and is designed to have a flat neutron efficiency, the average neutron energy can provide additional information about the reaction and be useful for efficiency correction in cases where one or several neutrons are emitted with energies outside the flat-efficiency region. Such additional information can be obtained from the ring-ratio method \citep{Berman1975}, discussed previously in the context of Figure~\ref{fig:simulated_ring_ratio}. A comparison between the \geant\ simulations and the source measurements is shown in Figure~\ref{fig:ring_ratio_wpube}.

\begin{figure}[ht!]
 \begin{center}
 \includegraphics[width=0.5\columnwidth]{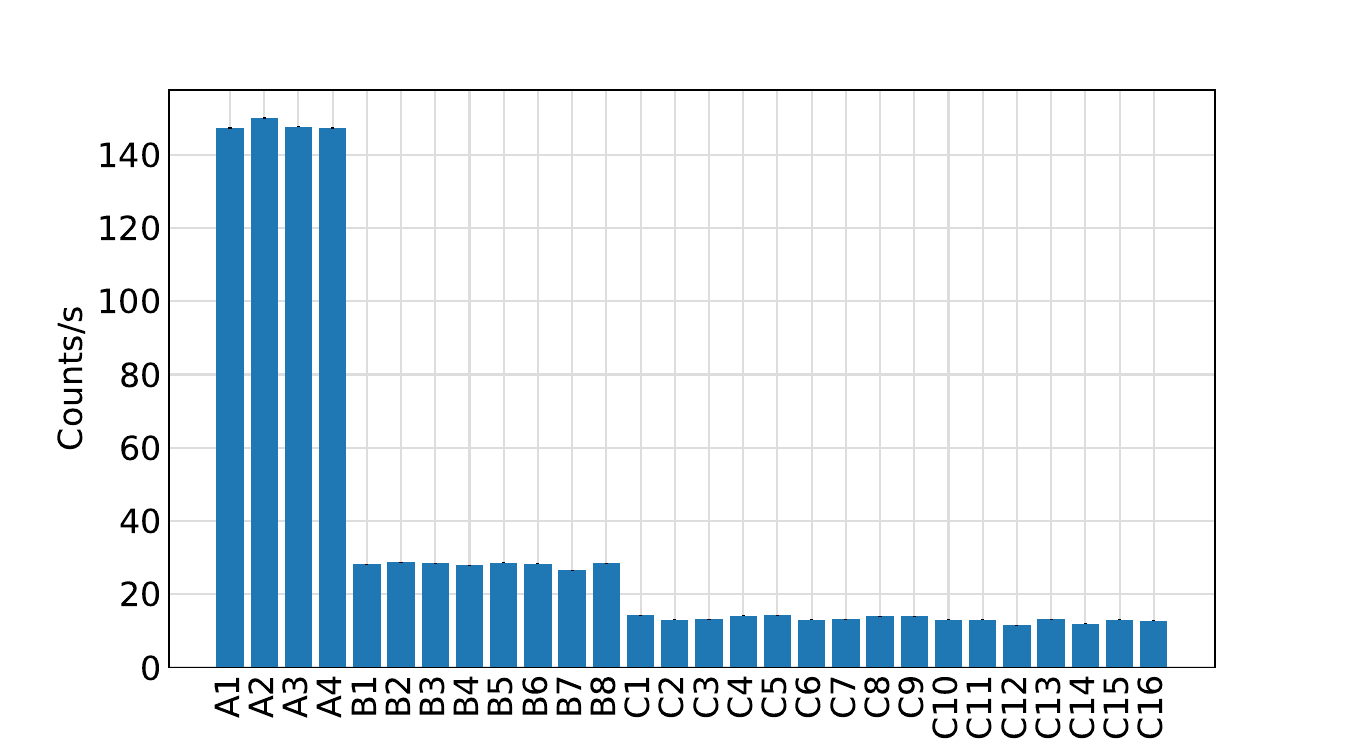}
 \includegraphics[width=0.5\columnwidth]{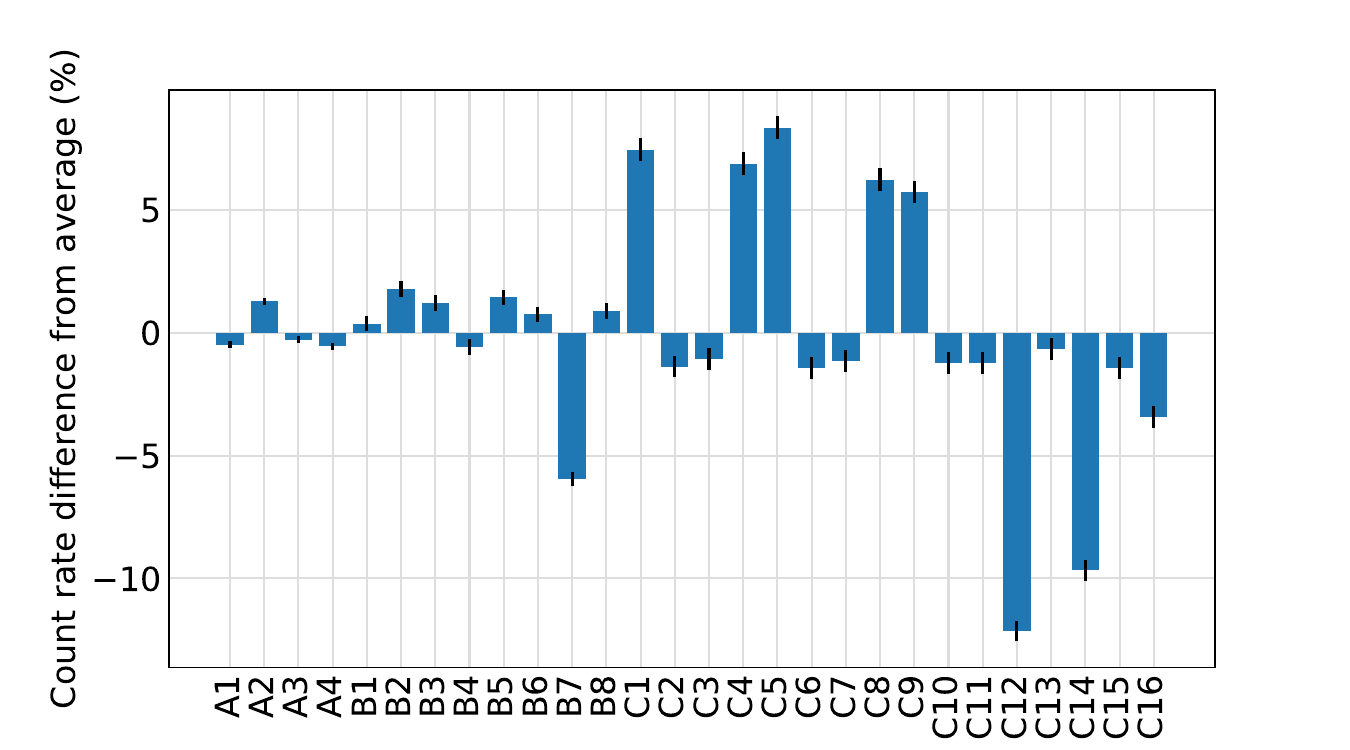}
\end{center}
\caption{Count rate in each neutron counter for the ${}^{252}$Cf source run (top) and deviation from the average between counters in the same ring (bottom).\label{fig:cf252_counts}}
\end{figure}

The ${}^{252}$Cf source measurements were performed in connection with the first in-beam experiment using \ac{ELIGANT-TN} at the 3MV~Tandetron accelerator setup, see Section~\ref{sec:inbeam}. The experiment was set up using the settings listed in Table~\ref{tab:det_table} and mounted on the beam line. Here, a source with a nominal activity of {\red{37.7(11)}}~kBq as of 2021-03-01 was used, with a calculated activity of {\red{24.6(7)}}~kBq at the time of measurement, 2022-09-22. With this source and in this configuration, the typical count rate in the individual counters is shown in Figure~\ref{fig:cf252_counts}.
The count rates were relatively constant within each ring, with a maximum deviation of approximately 10\% in the worst case, as shown in the bottom panel of Figure~\ref{fig:cf252_counts}. The measured efficiencies and ring ratios are shown in Figures~\ref{fig:efficiency_wpube} and \ref{fig:ring_ratio_wpube}, together with the measurements from the {\red{\ac{PuBe}}} source.

\begin{figure}[ht!]
 \begin{center}
 \includegraphics[width=0.5\columnwidth]{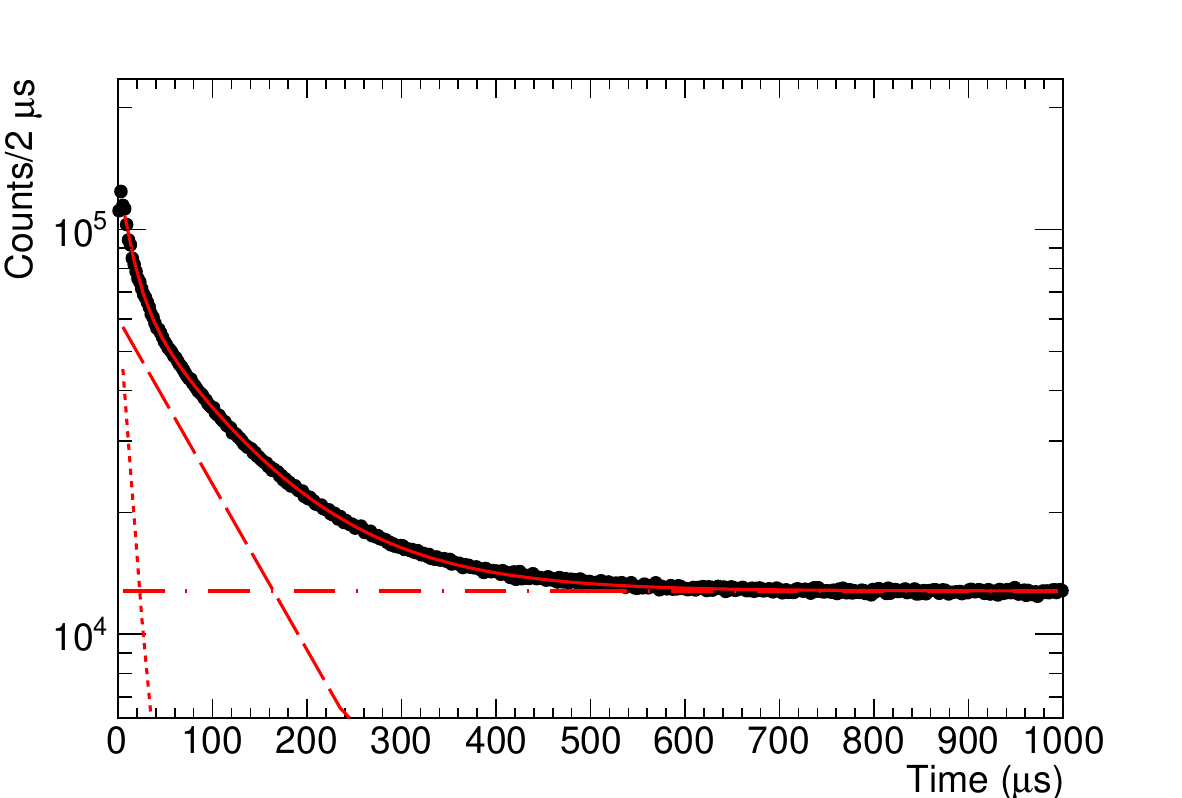}
\end{center}
\caption{Measured time spectrum of the detected neutron from a ${}^{252}$Cf neutron source. Two exponential decay components have been fitted to the simulated data with time constants {\red{$\tau_{1}=17.31(13)$~$\mu$s}} and {\red{$\tau_{2}=108.32(25)$~$\mu$s}} with about {\red{13.37(12)\%}} of the integrated counts in component 1 and about {\red{86.63(12)\%}} of the integrated counts in component 2.\label{fig:moderation_time}}
\end{figure}

In addition to verifying consistency between counters, the time distribution between detected neutrons was evaluated, as shown in Figure~\ref{fig:moderation_time}. 
This time approximately corresponds to the moderation time of neutrons with a ${}^{252}$Cf energy spectrum. However, as we did not have an independent fission trigger in this measurement, taking the first detected neutrons as a time reference will slightly distort this distribution. Still, the measured time-distribution agrees well with the \geant\ simulations, as consisting of two separate exponential components with time constants {\red{$\tau_{1}=17.31(13)$~$\mu$s}} and {\red{$\tau_{2}=108.32(25)$~$\mu$s}}, close to the \geant\ estimations of {\red{$\tau_{1}=10.27(35)$~$\mu$s}} and {\red{$\tau_{2}=108.2(9)$~$\mu$s}}, and the relative content of neutrons in the two integrals as {\red{13.37(12)\%}} and {\red{86.63(12)\%}}, respectively compared to {\red{25.0(5)}}\% and {\red{75.0(6)}}\% from \geant.

\section{{\red{In-beam measurements of $(\alpha,\mathrm{n})$ reactions}}\label{sec:inbeam}}

A test experiment to commission \ac{ELIGANT-TN} in-beam was performed at the 3~MV Tandetron accelerator at \ac{IFIN-HH} \cite{Burducea2015,Velisa2021} using a beam of $\alpha$ particles impinging on a ${}^{13}$C target with a thickness of 30~$\mu$m/cm$^{2}$. {\red{The target was placed in the centre of the \ac{ELIGANT-TN} detector array. The beam was extracted from a duoplasmatron ion source and the beam size was defined using two copper collimators with diameters of 5~mm and 3~mm, respectively.}} This induces the ${}^{13}\mathrm{C}(\alpha,\mathrm{n}){}^{16}\mathrm{O}$ reaction, providing monoenergetic neutrons at low energies and two possible neutron energies at high-beam energies. The digital electronics chain was used with the \ac{DELILA} \ac{DAQ} \cite{Aogaki2022,AogakiUnp} using CAEN V1725 digitisers. The data was collected in runs of around 5~min at the lower energies and around 15~min for the higher energies. A Faraday cup at the end of the beam line was used to monitor the beam current with an Ortec~439 current integrator. The current integrator was set to transmit a signal every $10^{-9}$~C. As this was mainly an in-beam commissioning of the setup, 
the simple Faraday cup setup was not optimised for high-precision absolute current measurements and, thus, only relative currents between different beam energies will be discussed in the following.
Thus, the absolute charge was not determined, and the cross-sections were normalised to previous measurements, with the primary goal being to reproduce the relative energy-dependency of the cross-section. The collected data were stored in the ROOT \cite{Brun1997,Antcheva2009} data format, with the energy obtained from \ac{DELILA}. For each run, the energy spectrum from each tube was integrated between a low-energy threshold and a high-energy threshold defined by the lowest wall-effect edge and the full-energy peak, see Figure~\ref{fig:he3_spectrum}.

\begin{figure}[ht!]
 \begin{center}
 \includegraphics[width=0.5\columnwidth]{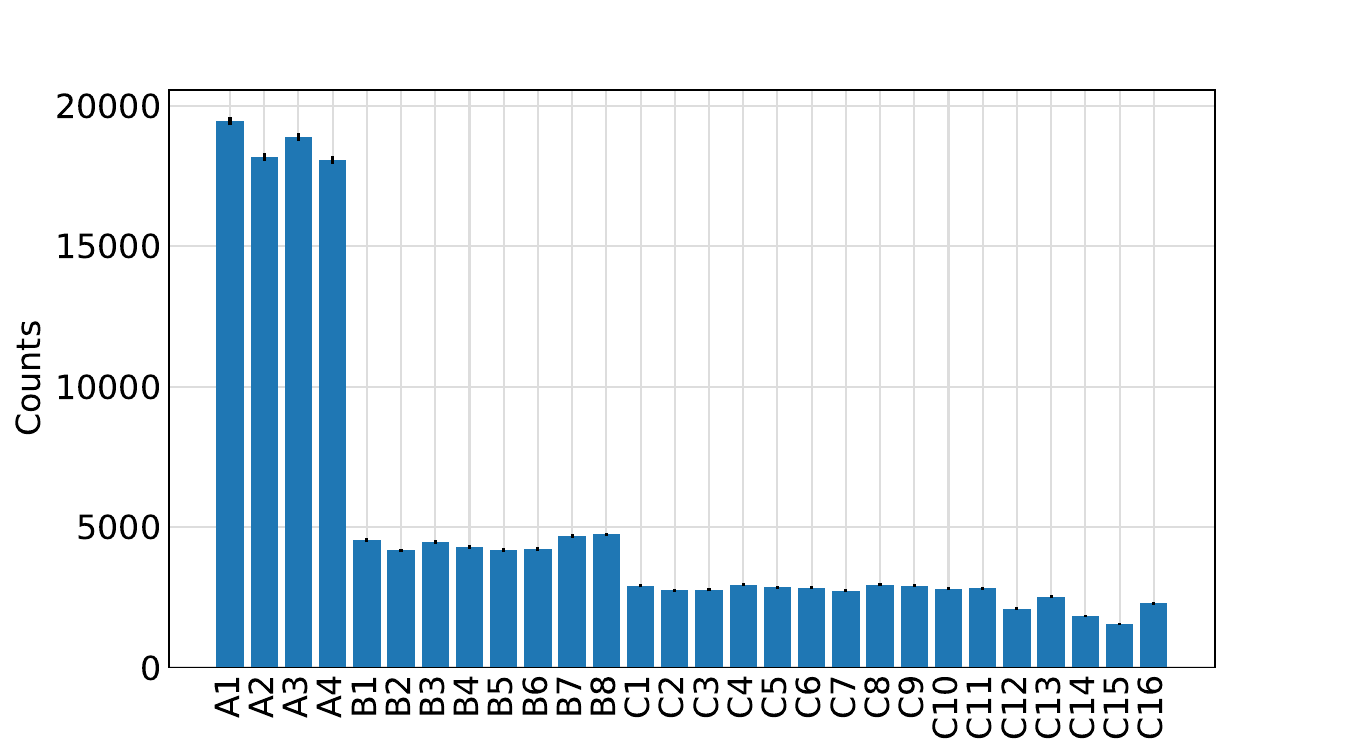}
 \includegraphics[width=0.5\columnwidth]{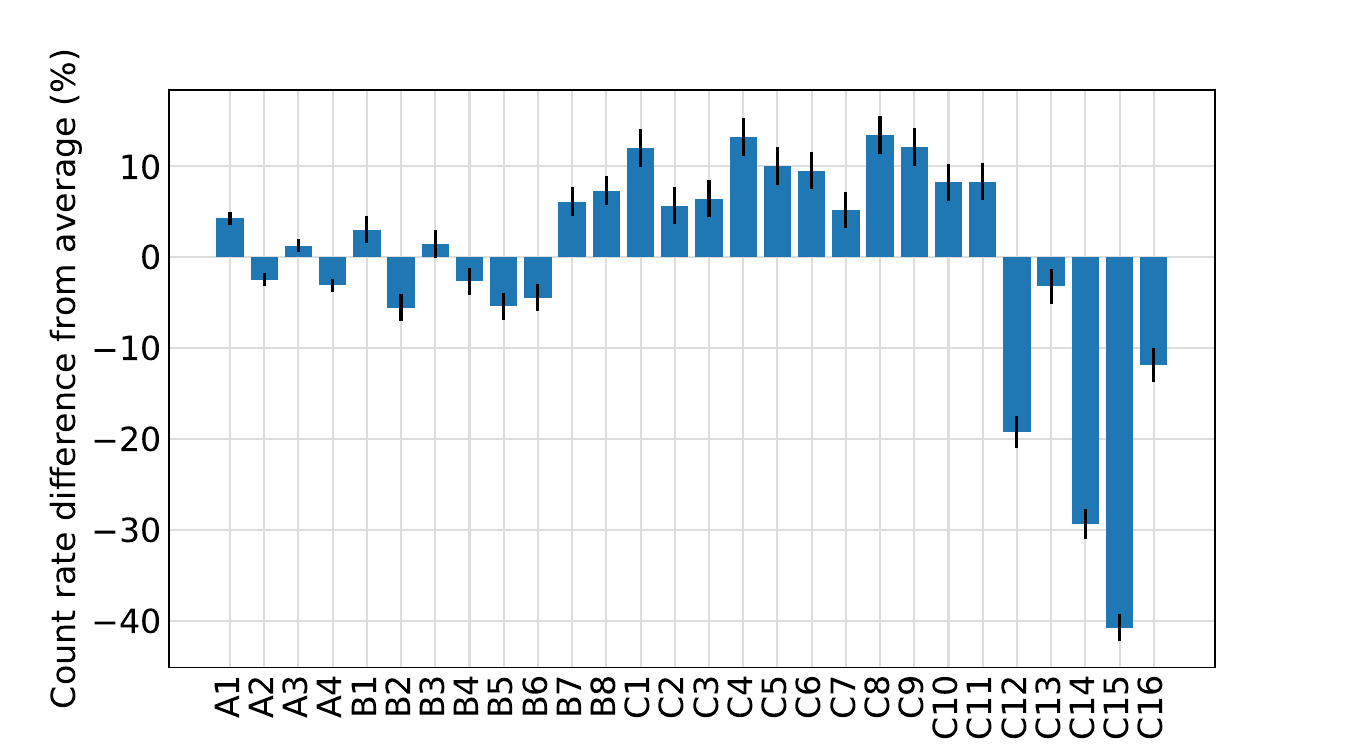}
\end{center}
\caption{Count rate in each neutron counter for the low beam-energy setting of the ${}^{13}$C runs between 1-1.15~MeV and 1.3-14~MeV.\label{fig:counts_c13}}
\end{figure}

The integrals of these spectra were saved individually for each neutron counter; see the example in Figure~\ref{fig:counts_c13}.
Ideally, each ring of \ac{ELIGANT-TN} should have the same count rate. However, this will not necessarily be the case in a real experimental environment where the count rate can differ due to \ac{HV} settings, electronics, beam properties, and similar. The relative count rate of each counter compared to the average value for each ring is also shown in Figure~\ref{fig:counts_c13}. While the inner two rings had a count-rate difference of less than 10\%, the outer ring had a significantly higher spread in count rate values, in some cases up to 40\% less count rate than the other counters in the ring. 

To calculate the actual energy of the $\alpha$ particle causing the reaction, the energy loss for the particle of a given energy was calculated over the entire target thickness, and it was assumed that the average energy loss was equal to half of this value. The beam energies from the logbook were corrected with these calculated average energy losses.

Following this, the \geant\ simulations of the efficiencies were extracted in the energy range 1~keV to 10~MeV. These efficiencies were extracted for the total efficiency and each of the individual rings of \ac{ELIGANT-TN}. These values were also used to extract simulated ring ratios for each possible combination of rings.

\begin{figure}[ht!]
 \begin{center}
 \includegraphics[width=0.5\columnwidth]{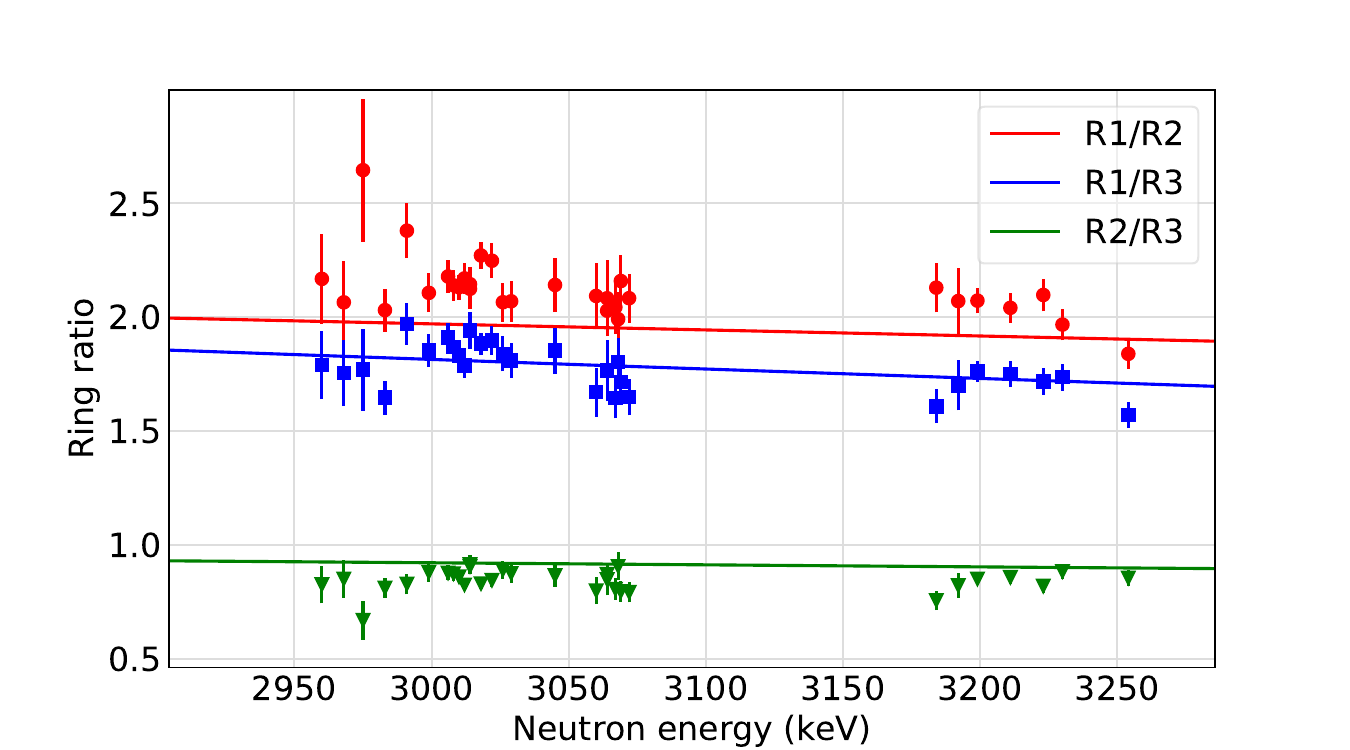}
\end{center}
\caption{Ring ratios for the different rings in ELIGANT-TN. The symbols correspond to the experimentally measured values and the lines to the \geant\ simulations.\label{fig:ring_ratios_lowe_13C}}
\end{figure}

For the low-energy beam settings, the neutron energies were estimated as {\red{$E_{\mathrm{n}}=E_{\mathrm{beam}}^{\mathrm{C.M.}}-\Delta E+2216$~keV}}, where $\Delta E$ correspond to the previously calculated energy loss of the beam in the target. The number of counts in the individual counters was combined into a number of counts in rings A, B, and C, corresponding to the inner, middle, and outer rings, respectively, and into the total number of counts for each of the rings combined. The standard deviation of the number of counts in the individual counters within a ring was extracted and added to the uncertainties as a systematic uncertainty, in addition to the statistical uncertainty from the number of counts in each counter, ring, or total. We can use this information to calculate the ring ratio at this lower beam energy and compare this to the values obtained from the \geant\ simulations. These ring ratios are shown in Figure~\ref{fig:ring_ratios_lowe_13C}.
While the agreement is good overall, we note that the inner ring, Ring~{\red{A}}, seems to have a higher count rate than expected, slightly distorting the measured ratios for Ring~{\red{A}} and Ring~{\red{B}}. However, the ring ratio for the outer rings shows a close agreement with \geant; we will mainly use this ratio in the later discussion.

In the higher energy range, a similar procedure was followed. As the reaction energy is above the first excited state in ${}^{16}$O, it is interesting to investigate the changes in the ring ratios. Here, we have remained with the two outer rings, assuming they are the most reliable and have used a simple two-component expression for the emitted neutrons as
\begin{equation}
    Y= Y_{\mathrm{g.s.}} + Y_{\mathrm{1st}}= \alpha Y + (1-\alpha)Y,
\end{equation}
where $Y_{\mathrm{g.s.}}$ is the number of neutrons populating the ground state in ${}^{16}$O, $Y_{\mathrm{1st}}$ is the number of neutrons populating the first excited state in ${}^{16}$O, $Y$ is the total number of neutrons detected, and $\alpha$ is the fraction of neutrons that populate the ground state. The measured ring ratio, say between Ring~{\red{B}} and Ring~{\red{C}}, for a single neutron energy, is simply
\begin{equation}
    R_{2/3} = \frac{\epsilon_{2}Y}{\epsilon_{3}Y}=\frac{\epsilon_{2}}{\epsilon_{3}}.
\end{equation}
However, the relation becomes more complex for two (or more) emitted neutrons. In the same case with two neutrons, it becomes
\begin{equation}
    R_{2/3} = \frac{\epsilon_{2}^{\mathrm{g.s.}}\alpha Y + \epsilon_{2}^{\mathrm{1st}}(1-\alpha)Y}{\epsilon_{3}^{\mathrm{g.s.}}\alpha Y + \epsilon_{3}^{\mathrm{1st}}(1-\alpha)Y}= \frac{\epsilon_{2}^{\mathrm{g.s.}}\alpha  + \epsilon_{2}^{\mathrm{1st}}(1-\alpha)}{\epsilon_{3}^{\mathrm{g.s.}}\alpha  + \epsilon_{3}^{\mathrm{1st}}(1-\alpha)},
\end{equation}
which needs to be solved for $\alpha$,
\begin{equation}
    \alpha = \frac{\frac{\epsilon_{3}^{\mathrm{1st}}}{\epsilon_{3}^{\mathrm{g.s.}}}\left(R_{2/3}^{\mathrm{1st}} -R_{2/3}\right)}{\left(R_{2/3} -R_{2/3}^{\mathrm{g.s.}}\right) + \frac{\epsilon_{3}^{\mathrm{1st}}}{\epsilon_{3}^{\mathrm{g.s.}}}\left(R_{2/3} + R_{2/3}^{\mathrm{1st}}\right)}.
\end{equation}

\begin{figure}[ht!]
 \begin{center}
 \includegraphics[width=0.5\columnwidth]{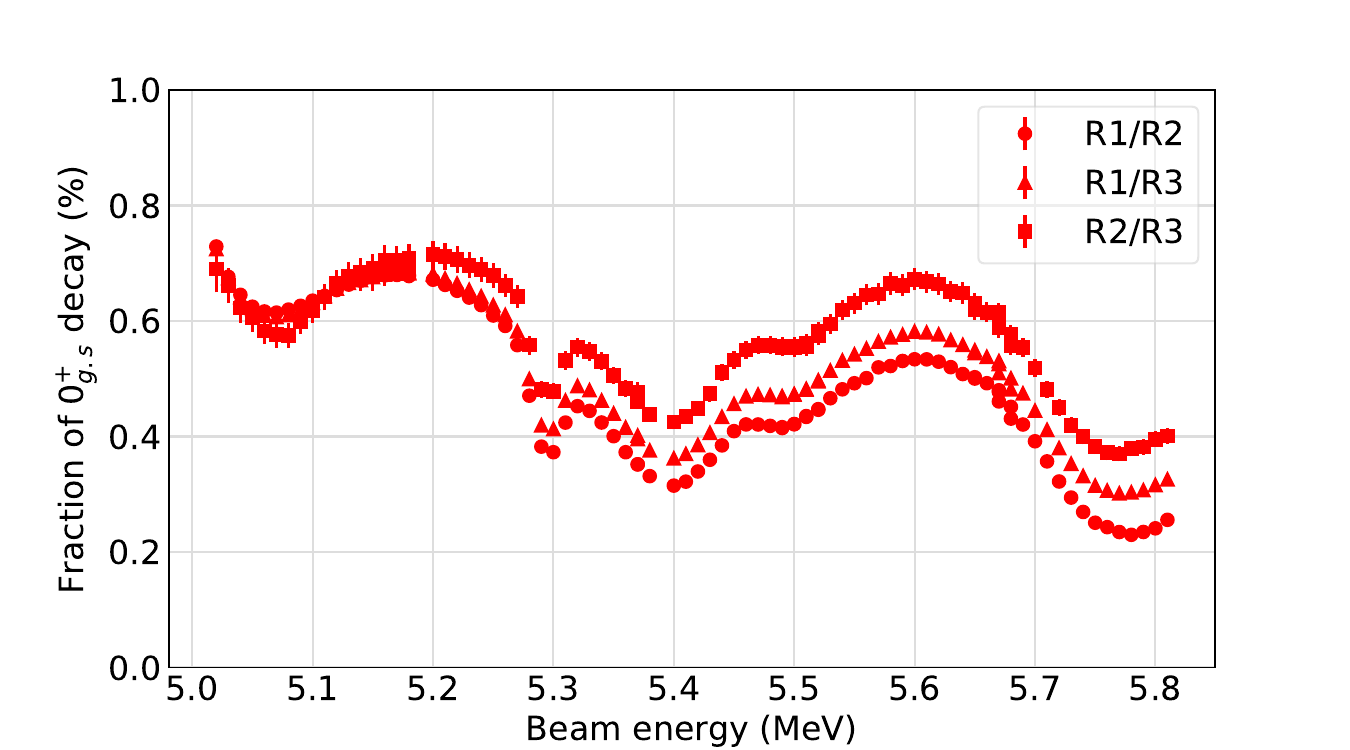}\\
 \includegraphics[width=0.5\columnwidth]{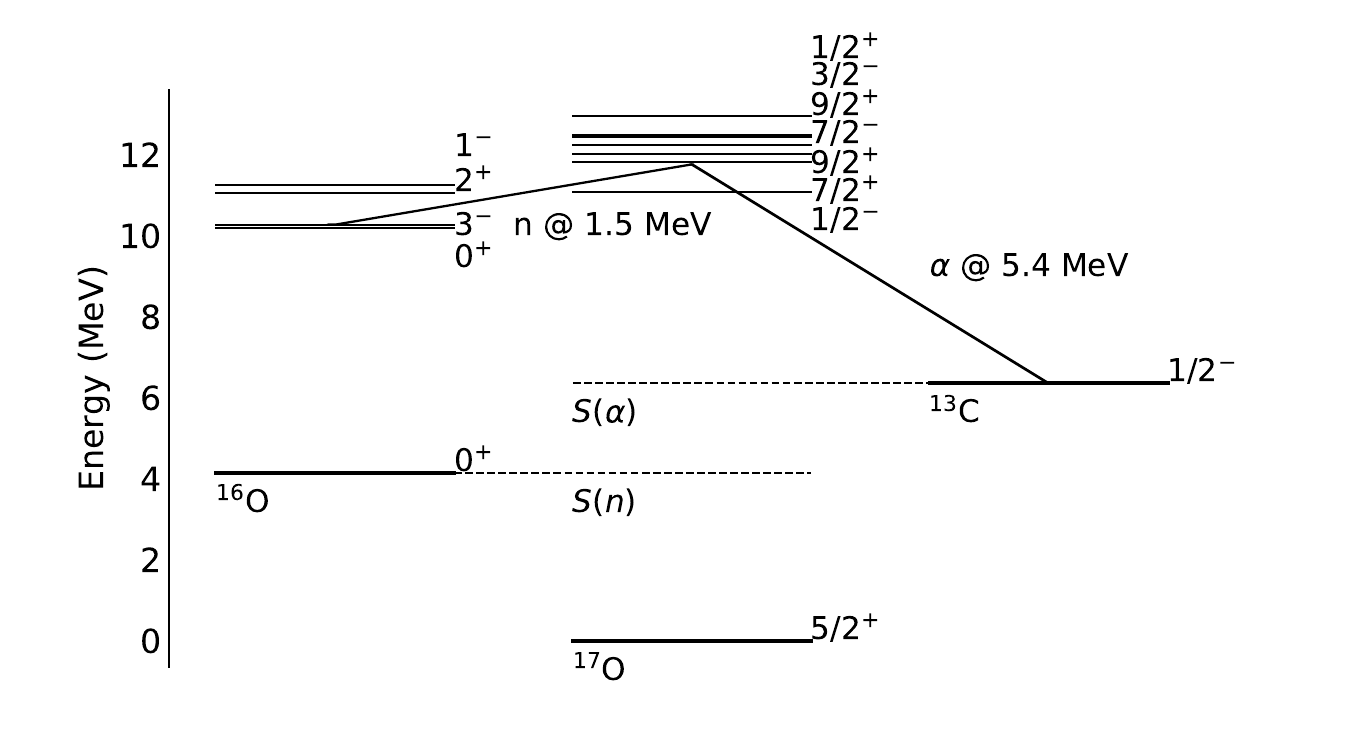}
\end{center}
\caption{(Top) Measured fraction of the neutron decay to the ground state in ${}^{16}$O relative to the fraction of decay to the first excited state. (Bottom) Schematic illustration of the neutron decay path to the lowest excited states.\label{fig:frac_gs_r2r3}}
\end{figure}

Given the experimental ring ratio between Ring~{\red{B}} and Ring~{\red{C}}, the value of $\alpha$ can be calculated from the \geant\ simulations, knowing that the neutron that populated the first excited state has an energy of $E_{\mathrm{n}}=E_{\mathrm{beam}}^{\mathrm{C.M.}}-\Delta E+2216-6049$~keV. From this relation, we can calculate $\alpha$ in the higher energy range with beam energies between 5-6~MeV, where we consider the beam-energy in the centre-of-mass frame, $E_{\mathrm{beam}}^{\mathrm{C.M.}}$. These fractions are shown in Figure~\ref{fig:frac_gs_r2r3}.
As expected, the decay fraction to the ground state decreases with increasing energy. In addition, we can see the appearance of some specific resonance-like structures, inverted, which suggests that these specific resonances are built upon excited states rather than on the ground state. 
The population of states in ${}^{17}$O was estimated using PACE4, which suggested a relatively broad distribution in angular momenta with the lowest angular momentum components as 8\% for $J=1/2$, 14\% for $J=3/2$, 15\% for $J=5/2$, 15\% for $J=7/2$, 13\% for $J=9/2$, and 10\% for $J=11/2$. As we expect the neutron to change the angular momentum with a relatively small amount when evaporated, we expect a preferred decay from $J=5/2,7/2,9/2$ states to the excited $3^{-}$ state in ${}^{16}$O with the decay to the $0^{+}$ ground state being suppressed, which could partially explain the relatively large branching to excited states.

\begin{figure}[ht!]
 \begin{center}
 \includegraphics[width=0.5\columnwidth]{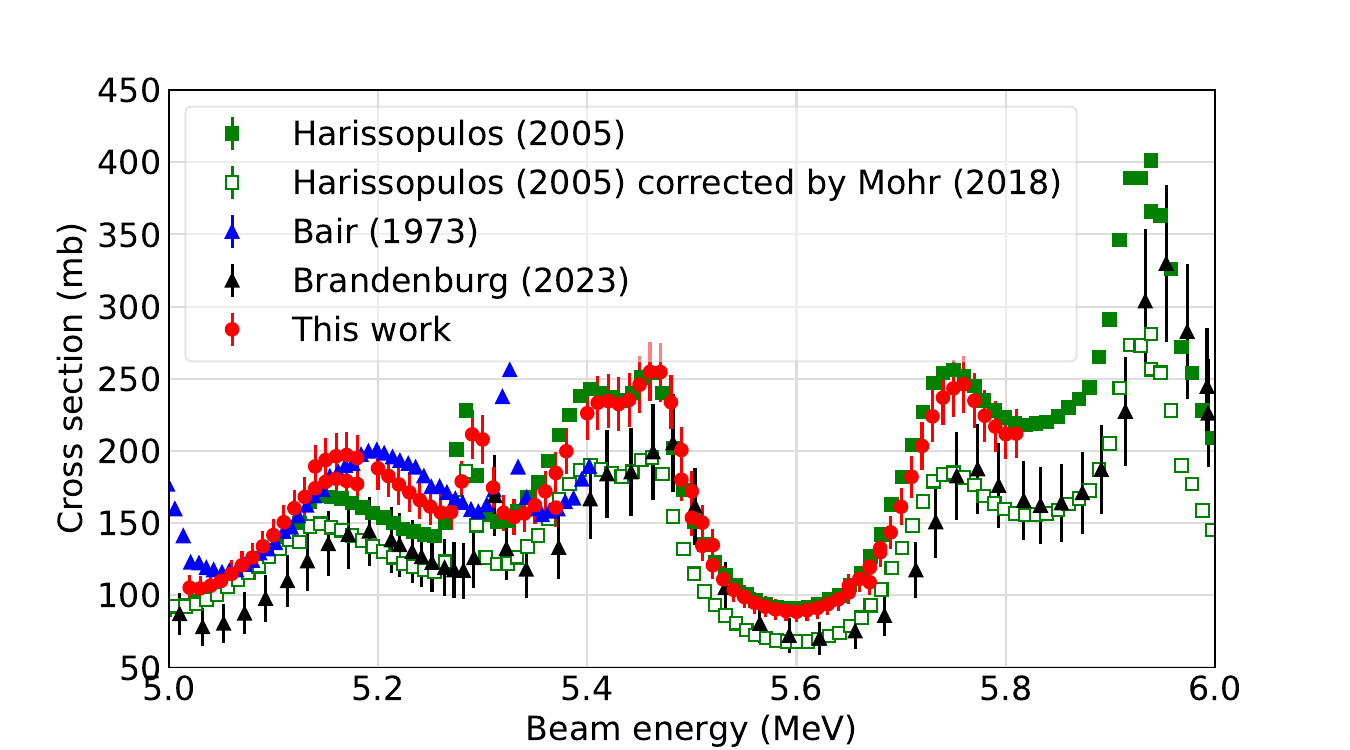}
\end{center}
\caption{Measured cross-sections in the high-energy range of ${}^{13}\mathrm{C}(\alpha,\mathrm{n}){}^{16}\mathrm{O}$ normalised to Harissopulos (2015) \cite{Harissopulos2005} and compared to Bair (1973) \cite{Bair1973}, {\red{Brandenburg (2023) \cite{Brandenburg2023}},} as well as the Harissopulos (2015) data corrected corrected by Mohr (2018) \cite{Mohr2018}.\label{fig:c13_highe_xsec}}
\end{figure}

Using these values of $\alpha$, we can deduce how we should weigh the efficiency of the setup during the measurements at these energies. As we are currently measuring cross-sections in the region where the \ac{ELIGANT-TN} flat efficiency is no longer valid, we must use the simulated efficiencies at two different energies. Thus, we modify the expression for the cross-section as
\begin{equation}
    \sigma[\mathrm{mb}] = \frac{R}{R_{\mathrm{cal}} e q } \frac{M}{z N_{\mathrm{A}}} \frac{10^{27}}{\alpha \epsilon(E_{\mathrm{n \to gs}})+(1-\alpha)\epsilon(E_{\mathrm{n \to 1st}})},
\end{equation}
where $R$ is the number of neutron counts divided by the number of signals from the current integrator, $R_{\mathrm{cal}}$ is the calibration value of the current integrator to translate the rate into elementary charges, $e$ is the elementary charge, $q$ is the charge state of the beam, $z$ is the target thickness in g/cm$^{2}$, $N_{\mathrm{A}}$ is Avogadro's constant, and $\epsilon$ is the efficiency calculated by \geant\ for each energy. With this expression, the measured cross sections in the high-energy range become as shown in Figure~\ref{fig:c13_highe_xsec}, compared with those from references \cite{Bair1973,Harissopulos2005,Brandenburg2023}. 

One problem with the data from Ref.~\cite{Harissopulos2005}, as pointed out by Peters and Mohr \cite{Peters2017,Mohr2018}, is how the experimental efficiency is treated. In particular, in Ref.~\cite{Harissopulos2005}, the neutrons are assumed to populate the ground state of ${}^{16}$O and the experimental data has been corrected accordingly, however it was argued \cite{Peters2017,Mohr2018} that this not necessarily had to be the case, which is also consistent with the ring ratio estimates shown here in Figure~\ref{fig:frac_gs_r2r3}. Instead, Mohr suggested a correction factor \cite{Mohr2018} based on estimation of the distribution of population of excited states from the \talys\ code \cite{Koning2008,Koning2012}. However, as Mohr correctly points out \cite{Mohr2018}, the validity of the statistical model approach for such light systems is not apparent, even if careful fine-tuning can reproduce some important features. In Figure~\ref{fig:c13_highe_xsec}, we also include the cross-section with the correction of Mohr \cite{Mohr2018}. What is curious to observe is that the data shown here, utilising the ring ratio from \ac{ELIGANT-TN}, still shows a relatively good agreement with the uncorrected data compared to the corrected data. 

\begin{figure}[ht!]
 \begin{center}
 \includegraphics[width=0.5\columnwidth]{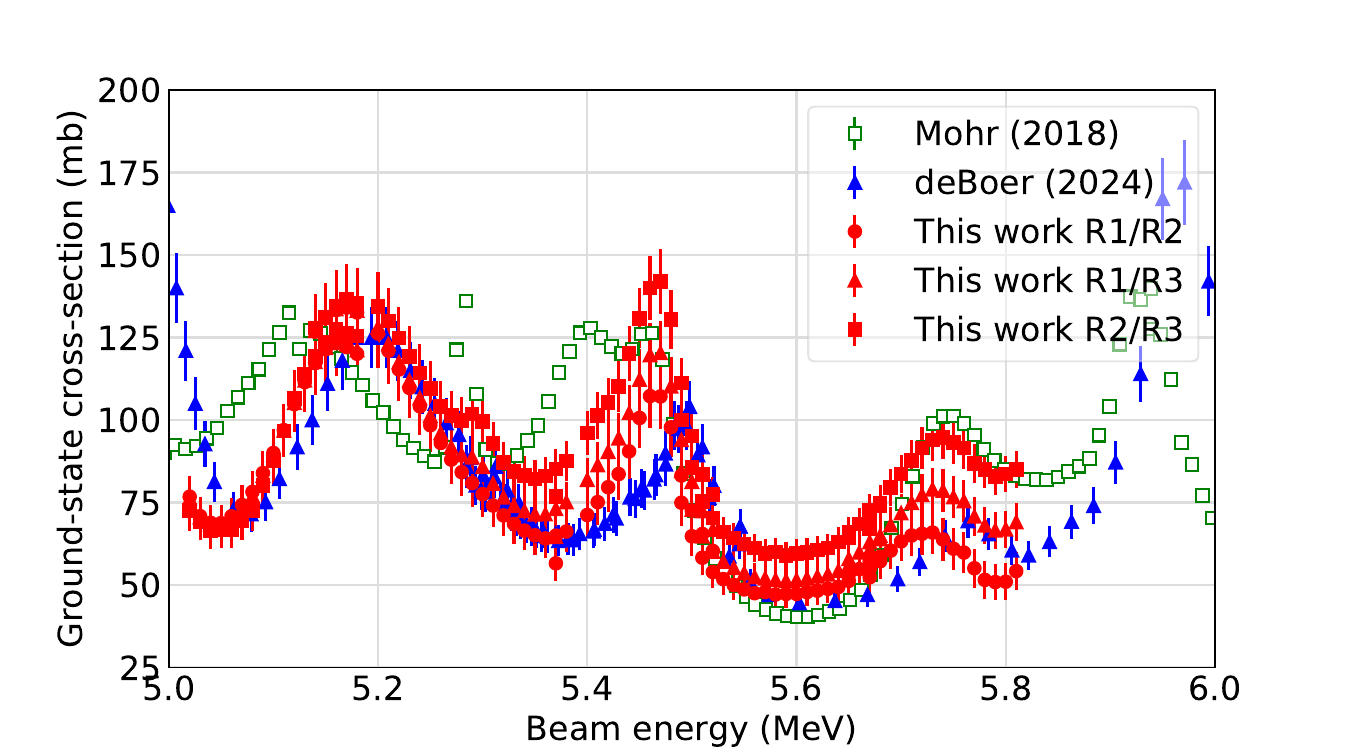}
\end{center}
\caption{Measured ground-state fraction of the cross-sections in the high-energy range of ${}^{13}\mathrm{C}(\alpha,\mathrm{n}_{0}){}^{16}\mathrm{O}$, with the total cross-section normalised to Harissopulos (2015) \cite{Harissopulos2005}, compared to the Harissopulos (2015) \cite{Harissopulos2005} data corrected corrected by Mohr (2018) \cite{Mohr2018} and the deBoer (2024) \cite{deBoer2024} data. {\red{The relative consistency between the exclusive ${}^{13}\mathrm{C}(\alpha,\mathrm{n}_{0}){}^{16}\mathrm{O}$ measurement by deBoer (2024) \cite{deBoer2024} and the inclusive measurement in this work shows that the ring-ratio method can be used with the ELIGANT-TN setup to separate different neutron energies, however, with significant uncertainties.}}\label{fig:c13_highe_xsec0}}
\end{figure}

In Reference~\cite{Gheorghe2024}, measurements of partial cross sections for neutron emission to low lying states in the residual based on the average neutron energy has been performed for ${}^{208}$Pb. In a similar manner, to test the validity of the ring-ratio approach to partial cross-sections we use the information on the estimated ground-state decay fraction from our commissioning measurement to compare with the recently published data \cite{deBoer2024} from the \ac{ODeSA}, consisting of nine deuterated scintillators mounted on a swing arm that could be rotated across a broad angular range to give explicit information about neutron decay energies and angular distributions, from an experiment at \ac{ORNL}. This measurement with an independent method is shown together with the corrected data from Ref.~\cite{Mohr2018}, and our data here utilising the ring-ratio method to estimate the fraction of ground-state decay, shown in Figure~\ref{fig:c13_highe_xsec0}.
While there is some disagreement, especially at the lowest and highest energies, the \ac{ELIGANT-TN} data and the \ac{ODeSA} data are relatively consistent within uncertainties, showing that the ring-ratio method indeed can be used in future experiments at \ac{ELI-NP} also for a more in-depth investigation of the neutron-emitting photonuclear reaction mechanisms. However, to fully resolve the situation in ${}^{13}$C, dedicated experiments optimised and carefully tuned for this specific problem should be performed.

\section{Summary and conclusions}

We have demonstrated the implementation of the ELIGANT-TN array, a component of the ELIGANT set of instruments for nuclear physics experiments above the neutron threshold, to be performed at ELI-NP. We have described the detector system, performed simulations using \geant\ and MCNP, performed commissioning tests using neutron sources, as well as an in-beam commissioning reproducing the ground-state cross-section fraction of the ${}^{13}\mathrm{C}(\alpha,\mathrm{n}){}^{16}\mathrm{O}$ reaction.

\section*{Author statement}
{\bf{PAS:}} Methodology, Project administration, Software, Formal analysis, Investigation, Writing - Original Draft, Writing - Review \& Editing, Visualization.
{\bf{DLB:}} Project administration.
{\bf{MC:}} Conceptualization, Resources, Writing - Original Draft, Visualization.
{\bf{DMF:}} Conceptualization, Methodology, Project administration, Resources. 
{\bf{IG:}} Conceptualization, Methodology.
{\bf{AK:}} Investigation.
{\bf{CM:}} Methodology, Software, Formal analysis, Writing - Original Draft.
{\bf{DT:}} Investigation.
{\bf{SA:}} Software.
{\bf{HTA:}} Investigation.
{\bf{LC:}} Investigation.
{\bf{DC:}} Investigation.
{\bf{GC:}} Conceptualization, Resources, Visualization.
{\bf{TG:}} Investigation.
{\bf{MK:}} Investigation.
{\bf{VL:}} Investigation.
{\bf{RR:}} Investigation.
{\bf{RFA:}} Investigation.
{\bf{MB:}} Investigation.
{\bf{RC:}} Investigation.
{\bf{AD:}} Investigation.
{\bf{DI:}} Investigation.
{\bf{DK:}} Investigation.
{\bf{SI:}} Investigation.
{\bf{KKH:}} Investigation.
{\bf{GL:}} Investigation.
{\bf{BM:}} Investigation.
{\bf{TP:}} Investigation.
{\bf{GVT:}} Investigation.

\section*{Acknowledgements}

The authors would like to acknowledge the support from the Extreme Light Infrastructure Nuclear Physics (ELI-NP) Phase II, a project co-financed by the Romanian Government and the European Union through the European Regional Development Fund - the Competitiveness Operational Programme (1/07.07.2016, COP, ID 1334) and the Romanian Ministry of Education and Research under research contract PN~23~21~01~06. PAS also acknowledges the support from the Romanian Ministry of Research, Innovation, and Digitalization via the Institute of Atomic Physics, M\u{a}gurele, Romania, contract number ELI-RO/RDI/2024-002. Experiments carried out at the 3~MV Tandetron\texttrademark\ accelerator from ``Horia Hulubei'' National Institute for Physics and Nuclear Engineering (IFIN-HH) were supported by Ministry of Education and Research, National Authority of Research, under the research contract PN~23~21~02~01 and by the Romanian Government Programme through the National Programme for Infrastructure of National Interest (IOSIN). We would like to thank N.~Safca, E.~Anghel, and D.~Stutman from the X-ray laboratory at ELI-NP/IFIN-HH for their support in the investigation of the internal structure of the tubes.




\bibliographystyle{elsarticle-num}






\acrodef{AGATA}{Advanced GAmma Tracking Array}
\acrodef{ALBA}{African LaBr Array}
\acrodef{AMD}{antisymmetrized molecular dynamics}
\acrodef{BGO}{bismuth germanate}
\acrodef{BNC}{Bayonet Neill-Concelman}
\acrodef{BRIKEN}{Beta-delayed neutrons at RIKEN}
\acrodef{BRUSLIB}{BRUSsels Nuclear LIBrary}
\acrodef{CAKE}{Coincidence Array for K600 Experiment}
\acrodef{CAD}{computer-aided design}
\acrodef{CCB}{Centrum Cyklotronowe Bronowice}
\acrodef{CFD}{constant-fraction discriminator}
\acrodef{CLHEP}{Class Library for High Energy Physics}
\acrodef{CMB}{cosmic microwave-background}
\acrodef{CRP}{Coordinated Research Project}
\acrodef{DAQ}{data acquisition}
\acrodef{DC}{direct current}
\acrodef{DELILA}{Digital ELI List-mode Acquisition}
\acrodef{DPP}{Digital Pulse-Processing}
\acrodef{DPP-PHA}{Digital Pulse Processing for Pulse Height Analysis}
\acrodef{DPP-PSD}{Digital Pulse Processing for Charge Integration and Pulse Shape Discrimination}
\acrodef{DSSSD}{double-sided silicon strip detector}
\acrodef{D-sub}{D-subminiature}
\acrodef{E8}{Experimental Area 8}
\acrodef{E9}{Experimental Area 9}
\acrodef{ECL}{emitter-coupled logic}
\acrodef{EDF}{Energy Density Functional}
\acrodef{ELI}{Extreme Light Infrastructure}
\acrodef{ELIADE}{ELI Array of DEtectors}
\acrodef{ELI-BIC}{ELI Bragg ionization chamber}
\acrodef{ELIGANT}{ELI Gamma Above Neutron Threshold}
\acrodef{ELIGANT-GG}{ELIGANT Gamma Gamma}
\acrodef{ELIGANT-GN}{ELIGANT Gamma Neutron}
\acrodef{ELIGANT-TN}{ELIGANT Thermal Neutron}
\acrodef{ELI-GBS}{ELI $\gamma$-ray Beam System}
\acrodef{ELI-NP}{Extreme Light Infrastructure -- Nuclear Physics}
\acrodef{ELISSA}{ELI Silicon Strip Array}
\acrodef{EWSR}{energy-weighted sum rule}
\acrodef{FADC}{flash ADC}
\acrodef{FASTER}{Fast Acquisition System for nuclEar Research}
\acrodef{FATIMA}{FAst TIming Array}
\acrodef{FFT}{fast Fourier transform}
\acrodef{FOM}{figure-of-merit}
\acrodef{FREYA}{Fission Reaction Event Yield Algorithm}
\acrodef{FWHM}{full width at half maximum}
\acrodef{GDR}{giant dipole-resonance}
\acrodef{GECO2020}{GEneral COntrol Software}
\acrodef{GLO}{Generalized Lorentzian}
\acrodef{GRAF}{Grand RAiden Forward}
\acrodef{GROOT}{\geant\ and ROOT Object-Oriented Toolkit}
\acrodef{gSF}[$\gamma$SF]{$\gamma$-ray strength function}
\acrodef{GSI}{Gesellschaft f\"{u}r Schwerionenforschung}
\acrodef{HDPE}{high-density polyethylene}
\acrodef{HeBGB}{${}^{3}$He BF$_{3}$ Giant Barrel}
\acrodef{HF}{Hartree-Fock}
\acrodef{HFB}{Hartree-Fock-Bogolyubov}
\acrodef{HIgS}[HI$\gamma$S]{High-Intensity $\gamma$-ray Source}
\acrodef{HPGe}{high-purity germanium}
\acrodef{HPLS}{high-power laser system}
\acrodef{HV}{high voltage}
\acrodef{IAEA}{International Atomic Energy Agency}
\acrodef{IFIN-HH}{Horia Hulubei Institute for Physics and Nuclear Engineering}
\acrodef{iThemba LABS}{iThemba Laboratory for Accelerator Based Sciences}
\acrodef{JINR}{Joint Institute for Nuclear Research}
\acrodef{KMF}{Kadmenskii-Markushev-Furman}
\acrodef{KRATTA}{Krak\'{o}w Triple Telescope Array}
\acrodef{LANL}{Los Alamos National Laboratory}
\acrodef{LCS}{laser Compton backscattering}
\acrodef{LED}{leading-edge discriminator}
\acrodef{LEDR}{low-energy electric dipole response}
\acrodef{LLNL}{Lawrence Livermore National Laboratory}
\acrodef{MDA}{multipole decomposition analysis}
\acrodef{MDR}{magnetic dipole-resonance}
\acrodef{MCA}{multi-channel analyzer}
\acrodef{MCNP}{Monte Carlo N-Particle Transport Code}
\acrodef{MCX}{micro coaxial connector}
\acrodef{MIDAS}{Multi Instance Data Acquisition System}
\acrodef{MONSTER}{MOdular Neutron time-of-flight SpectromeTER}
\acrodef{MWDC}{Multi-Wire Drift Chambers}
\acrodef{NDF}{number of degrees of freedom}
\acrodef{NEDA}{NEutron Detector Array}
\acrodef{NIM}{Nuclear Instrumentation Module}
\acrodef{NLD}{nuclear level density}
\acrodefplural{NLD}[NLDs]{nuclear level densities}
\acrodef{NNDC}{National Nuclear Data Center}
\acrodef{NRF}{nuclear resonance fluorescence}
\acrodef{OCL}{Oslo Cyclotron Laboratory}
\acrodef{ODeSA}{Oak Ridge National Laboratory Deuterated Spectroscopic Array}
\acrodef{ORNL}{Oak Ridge National Laboratory}
\acrodef{OSCAR}{Oslo Scintillator Array}
\acrodef{PANDORA}{Photo-Absorption of Nuclei and Decay Observables for Reactions in Astrophysics}
\acrodef{PARIS}{Photon Array for Studies with Radioactive Ion and Stable Beams}
\acrodef{PCB}{printed circuit board}
\acrodef{PCIe}{Peripheral Component Interconnect Express}
\acrodef{PDR}{pygmy dipole resonance}
\acrodef{PLL}{phase-locked loop}
\acrodef{PMT}{photomultiplier tube}
\acrodef{PRD}{prompt-response distribution}
\acrodef{PSD}{pulse-shape discrimination}
\acrodef{PuBe}[{${}^{239}$Pu/Be}]{plutonium-239/beryllium}
\acrodef{QED}{quantum electrodynamics}
\acrodef{QPM}{quasiparticle-phonon model}
\acrodef{QRPA}{Quasiparticle Random Phase Approximation}
\acrodef{RC}{resistor-capacitor}
\acrodef{RCNP}{Research Center for Nuclear Physics}
\acrodef{RF}{radio frequency}
\acrodef{RIPL}{Reference Input Parameter Library}
\acrodef{RMS}{root-mean-square}
\acrodef{ROSPHERE}{ROmanian array for SPectroscopy in HEavy ion REactions}
\acrodef{RQRPA}{Relativistic Quasiparticle Random Phase Approximation}
\acrodef{RQTBA}{Relativistic Quasiparticle Time-Blocking Approximation}
\acrodef{SHV}{safe high voltage}
\acrodef{SAKRA}{Si Array developed by Kyoto and osaka for Research into Alpha cluster states}
\acrodef{SLEGS}{Shanghai Laser Electron Gamma Source}
\acrodef{SMLO}{simplified version of the modified Lorentzian}
\acrodef{SORCERER}{SOlaR CElls for Reaction Experiments at ROSPHERE}
\acrodef{SSC}{Separated Sector Cyclotron}
\acrodef{TCP/IP}{Transmission Control Protocol/Internet Protocol}
\acrodef{TDR}{technical design report}
\acrodef{TRK}{Thomas-Reiche-Kuhn}
\acrodef{TOF}{time-of-flight}
\acrodef{TUD}[TU Darmstadt]{Technische Universität Darmstadt}
\acrodef{TUNL}{Triangle Universities Nuclear Laboratory}
\acrodef{UHECR}{ultra-high energy cosmic ray}
\acrodef{UPC}{Universitat Polit\`{e}cnica de Catalunya}
\acrodef{USB}{Universal Serial Bus}
\acrodef{VECC}{Variable Energy Cyclotron Centre}
\acrodef{VEGA}{Variable Energy Gamma-ray}
\acrodef{VME}{Versa Module Europa}

\acrodef{F4}{Average Cell Flux}
\end{document}